\documentclass[preprint]{aastex}

\shorttitle{Structure of young M31 clusters}
\shortauthors{Barmby et al.}

\begin{document}

\title{An HST/WFPC2 Survey of Bright Young Clusters in M31 III. Structural Parameters%
\footnote{Based on observations made with the NASA/ESA 
Hubble Space Telescope, obtained at the Space Telescope Science Institute, 
which is operated by the Association of Universities for Research in 
Astronomy, Inc., under NASA contract NAS 5-26555. 
These observations are associated with program GO-10818 (PI J. Cohen) and GO-8296 (PI P. Hodge).}
}

\author{P. Barmby\altaffilmark{1},
S. Perina\altaffilmark{2,3}, 
M. Bellazzini\altaffilmark{2,3},
J. G. Cohen\altaffilmark{4}, 
P.W. Hodge\altaffilmark{5}, 
J.P. Huchra\altaffilmark{6},
M. Kissler-Patig\altaffilmark{7},
T.H. Puzia\altaffilmark{8},
J. Strader\altaffilmark{6,9}
}

\altaffiltext{1}{Department of Physics \& Astronomy, University
of Western Ontario, London, ON N6A 3K7, Canada; e-mail: pbarmby@uwo.ca}
\altaffiltext{2}{INAF - Osservatorio Astronomico di Bologna, via Ranzani 1, 40127 Bologna, Italy}
\altaffiltext{3}{Universit\`a di Bologna, Dipartimento di Astronomia, via Ranzani 1, 40127 Bologna, Italy}
\altaffiltext{4}{Palomar Observatory, Mail Stop 105-24, California Institute of Technology, Pasadena, CA 91125}
\altaffiltext{5}{Department of Astronomy, University of Washington, Seattle, WA 98195, USA}
\altaffiltext{6}{Harvard-Smithsonian Center for Astrophysics, Cambridge, MA 02138 USA}
\altaffiltext{7}{European Southern Observatory, Karl-Schwarzschild-Strasse 2, 85748 Garching bei M\"unchen, Germany}
\altaffiltext{8}{Herzberg Institute of Astrophysics, 5071 West Saanich Road, Victoria, BC V9E 2E7, Canada}
\altaffiltext{9}{Hubble Fellow}

\begin{abstract}
Surface brightness profiles for 23 M31 star clusters 
were measured using images from the Wide Field Planetary Camera 2 
on the {\it Hubble Space Telescope}, and fit to two types of models to 
determine the clusters' structural properties. The clusters are
primarily young ($\sim 10^8$~yr) and massive ($\sim10^{4.5}$~M$_\sun$),
with median half-light radius 7~pc and dissolution times of a few Gyr.
The properties of the M31 clusters 
are comparable to those of clusters of similar age in the Magellanic Clouds. 
Simulated star clusters are used to derive a conversion from statistical
measures of cluster size to half-light radius so that the extragalactic
clusters can be compared to young massive clusters in the Milky Way. 
All three sets of star clusters fall approximately on the same age-size relation.
The young M31
clusters are expected to dissolve within a few Gyr and will not survive to become 
old, globular clusters. However, they do appear to follow the same fundamental
plane relations as old clusters; if confirmed with velocity dispersion measurements,
this would be a strong indication that the star cluster fundamental plane
reflects universal cluster formation conditions.
\end{abstract}


\section{Introduction}
The spatial distribution of stars within a star cluster is an important 
indicator of the cluster's dynamical state, and the structural parameters 
(e.g. core, half-light, and tidal radii; central surface brightness, and concentration) 
indicate on what timescales the cluster is 'bound' to dissolve.  
The work of  \citet{spitzer87} showed that core collapse is an inevitable 
part of cluster dynamical evolution. \citet{djorgovski86} were among the first
to determine the fraction of core-collapsed Milky Way globular clusters (GCs), while 
\citet{djorgovski94} examined a large sample
of Milky Way clusters and defined the `fundamental plane', showing that surface
brightness profiles of Galactic GCs were well-described by only a few parameters.
\citet{meylan87} surveyed GCs in the LMC and SMC for core collapse and found that
only a handful of clusters were core-collapse candidates; they suggested that environmental
or age effects were responsible for the difference with Milky Way globulars.

A few spatially-resolved studies of GCs beyond the Magellanic Clouds were done with ground-based data.
\citet{racine91} and \citet{racine92} used high-resolution imaging to distinguish M31 GC candidates from
background galaxies, and \citet{cohen91} determined the tidal radii of 30 M31 halo GCs,
finding them to be similar to Milky Way GCs. However, detailed studies of the structures of M31 GCs
awaited the angular resolution of the \textit{Hubble Space Telescope}.
The first work on M31 GCs by \citet{bendinelli93} and \citet{fusipecci94} was followed by numerous others including
\citet{rich96}, \citet{grillmair96},
\citet{holland97}, and \citet{barmby02,barmby07}.  Clusters 
 in Local Group galaxies are near the limit for
resolution into individual stars by the {\it Hubble Space Telescope}
(HST), although some structural information such as half-light radii
can be recovered for clusters in more distant galaxies \citep[e.g.,][]{hasegan05}. 
Conclusions of the studies of extragalactic globulars include the dependence of cluster size on galactocentric
radius,  first pointed out for the Milky Way by \citet{djorgovski94} and \citet{vdb94};
a possible difference between sizes of clusters in different metallicity groups \citep[for
a detailed discussion see][]{jordan04};
and a recognition that globular clusters in a variety of environments appear to lie on the same
fundamental plane \citep{barmby07}. 

Structural studies of younger star clusters present more difficulties. 
Open clusters (OCs) in the Milky Way are generally much less massive than 
globular clusters. As viewed from our location in the Milky Way, they are embedded within
the disk, so that the cluster is easily lost against the much more numerous
field stars, and determining stellar membership
in these less-concentrated objects is not straightforward.
Comprehensive studies of Milky Way open clusters are relatively recent:
\citet{kharchenko05} and follow-up work \citep{schilbach06, piskunov07,piskunov08}
measured a variety of radii (core, corona, tidal) for several hundred clusters
and found their masses to be in the range 50--1000 M$_\sun$.
\citet{bonatto05} analyzed 
in more detail a much smaller number of Milky Way open clusters,
finding that the cluster size increased with both age and Galactocentric distance.
These authors also found that their sample of clusters showed evidence for an
 `open cluster fundamental plane.'

Milky Way open clusters are not the only known population of young star clusters,
and possibly not even the best one to study. The Galactic OCs
cover a limited range in age and mass and their census  is suspected
to be far from complete because of extinction in the Galactic plane.
The Magellanic Clouds (MCs) have many young star clusters,
recently cataloged by \citet{bica08}. The brighter MC clusters were studied in a pioneering work by \citet{elson87}.
These authors analyzed the radial profiles of  10 clusters and found them to be
better-fit by `power-law' profiles of the form $I(R) \propto [1+(R/r_0)^2]^{-(\gamma-1)/2}$
than by the \citet{king66} models conventionally used to fit  globular cluster profiles.
\citet{mclaughlin05} re-analyzed a large set of MC cluster data and
found the situtation to be somewhat more complex. Those authors argued
that the extended envelopes characteristic of the power-law profiles are a generic
feature of many young {\it and old\/} star clusters and that
``the development of a physically motivated model accounting for this \dots 
could lend substantial new insight into questions of cluster formation and evolution.''

Outside the Milky Way, many galaxies are found to have `young massive clusters'
\citep[YMCs;][]{holtzman92,whitmore95}. These clusters have ages up to a few Gyr \citep{brodie98}
and masses comparable to globular clusters \citep{larsen99}.
Studies of YMC structures show correlations of power-law slope $\gamma$ with 
age \citep{larsen04}, core radius with age \citep{mackey03}, and mass of the brightest
cluster with galaxy star formation rate \citep{weidner04}. 
As of yet there is no comprehensive study of star cluster structures over the full age 
and mass ranges seen in nearby galaxies.
M31 is now recognized to also have a large population of young star clusters
\citep{fusipecci05,caldwell09}, although their relationship to both the YMCs 
and globular clusters is not well-understood. The purpose of this paper is to carry out an
initial study of the structural properties of some young M31 clusters. We analyze a sample of 23
clusters using data from the Wide Field Planetary Camera 2 (WFPC2) 
onboard the {\it Hubble Space Telescope}; extensive analysis of `artificial clusters'
(see Appendix) informs our analysis procedures.
Throughout this work we assume a distance to M31 of 783~kpc \citep{stanek98},
for which 1\arcsec\ corresponds to 3.797~pc. All magnitudes are in the Vega system,
and cluster names use the convention of the Revised Bologna Catalog \citep{rbc04};%
\footnote{Online version at {\url http://www.bo.astro.it/M31}}
see that work for cluster coordinates and other properties.

\section{Data and analysis methods}

\subsection{Cluster sample}

The study of star clusters in M31 has a long history dating back to at least 
\citet{hubble32}, so any attempt to assemble a sample of young massive clusters necessarily
draws on many previous works. While a number of studies of the \textit{globular} 
cluster system have noted the presence of possible young clusters in M31 
\citep{barmby00,williams01},
the first comprehensive list of such objects was assembled by \citet{fusipecci05}, who
called them `blue luminous compact clusters', or BLCCs. 
\citet{krienke07,krienke08} and \citet{hodge09}
searched for M31 `disk clusters' in archival HST imaging data, and \citet{caldwell09}
presented a comprehensive list of nearly 150 young cluster candidates from a spectroscopic survey.
\citet{caldwell09} noted that the handful of their young clusters with measured structural
properties \citep[from][]{barmby07} covered a wide range in parameter space.
The HST resolved-star study of four `massive and compact young star clusters'  by \citet{williams01} 
(program GO-8296) did not include an analysis of the objects' structural properties.

The main sample of clusters studied here is described in detail by the companion papers by
\citet{perina09,perina10}.
The present project began with an interest in confirming the results of \citet{cohen05}
who used adaptive optics imaging to show that some of the clusters proposed as
young were in fact asterisms 
(but see the contrary view of \citealt{caldwell09} and the discussion in \citealt{perina09}).
HST program GO-10818 was aimed at imaging all of the `class A' clusters proposed
by \citep{fusipecci05} which did not already have HST imaging, a total of 21 objects.
In the course of the program we found that two clusters in the candidate list were in 
fact the same object \citep{perina09}, and the object NB67 was a star, so the program contains
19 objects. \citet{perina10} showed that 16 of the clusters are young, with ages $<1$~Gyr,
and five (B083, B222,  B347, B374, and NB16) are in fact intermediate-aged or old \citep[see also][]{caldwell09}.
We retain these five clusters in our sample but show them with different symbols in the analysis.
We augmented the GO-10818 data with archival data on the four clusters
studied by \citet{williams01} to bring the total number of clusters to 23.
HST archival data exists for additional clusters but in the interests
of dealing with a mostly-homogeneous dataset we restricted the sample to
only the GO-10818 and GO-8296 clusters. Three of the clusters in
the latter dataset had structural parameters reported in \citet{barmby02};
here we re-analyze them in a manner consistent with the other clusters.
Except for B083 and B347, all of the clusters  are projected against
the M31 disk \citep[see Fig.~1 of][]{perina10}.

\subsection{Data reduction and surface brightness profiles}

The GO-10818 program was originally intended to be carried out with the Advanced Camera for Surveys
(ACS), but because that instrument failed, the images were obtained instead with 
the Wide-Field Planetary Camera 2 (WFPC2). All objects were observed with two 400-s dithered
images in each of 2 filters: F450W and F814W \citep[for further detail, and an example
of the CMD analysis, see][]{perina09}. 
The GO-8296 program was also carried out with WFPC2 and involved
two 800-s images in F439W and two 600-s images in F555W (as well as
longer images in F336W which are not used here). The target clusters were
on the PC chip in all cases, and only data from that chip is used in the present
analysis. Table~\ref{tab:hstdat} summarizes the datasets together with
other pertinent information about the clusters.

The multiple images were combined with the
STScI Multidrizzle software, using the `recipes' provided on the drizzle webpage.
The  pixel scale of the resulting images was 0.0455\arcsec,
or 0.172~pc at the M31 distance.
While correcting for Charge Transfer Efficiency losses would be 
desirable, there is currently no prescription
available for correcting surface photometry of extended objects so 
no correction has been made in the present analysis.
Although M31 star clusters are relatively large (a few arcsec) compared to the HST optical point-spread
function (PSF), convolving model profiles with the PSF prior to
comparison with the data should improve the
accuracy of measurements of the cluster cores.
Model PSFs were generated for the relevant filters
at the camera center using TinyTim. The clusters are small compared to
the camera field-of-view, and PSF variation over the cluster extent is negligible.

Transforming instrumental magnitudes to calibrated surface brightness was
done following the prescription in \citet{barmby07}. Image counts
were first multiplied by the inverse square of the pixel scale to
give counts $C$ in units of~s$^{-1}$~arcsec$^{-2}$. These can be transformed to 
magnitudes~arcsec$^{-2}$ through $\mu = Z-2.5\log(C)$, where $Z$
is the instrument zeropoint. They can also be transformed to intensity $I$ in 
$L_\sun$~pc$^{-2}$ through $I = 10^{0.4(Z^\prime-Z)}C$.
(Independent of the instrument used, $Z^\prime=(m-M)_{\rm M31}+M_\sun+5\log(\beta) = 21.5715+ M_\sun$
where $\beta$ is the number of arcsec corresponding to 1~pc; $\beta=0.2644$ at the
assumed distance of M31.)
The zeropoints used come from the respective instrument handbooks;
the solar magnitudes are from calculations by C. Willmer\footnote{\url{http://www.ucolick.org/~cnaw/sun.html}}. 
All are listed in Table~\ref{tab:caldat} for reference.

Studies of surface brightness profiles of  Local Group star clusters 
are in a somewhat different regime from either Galactic clusters
or clusters in more distant galaxies.  Local Group star clusters are
resolved into stars in their outer regions but not in their cores. They differ
from galaxies with comparable angular sizes 
($\lesssim 10$ arcsec for M31 and M33 clusters) in that the galaxies are composed of 
many more stars and have much smoother light distributions.
To better understand the limitations of our analysis, we simulated
artificial star clusters, measured their surface brightness profiles, and
fit those profiles to models: these simulations are described in
Appendix~\ref{sec:artclust}.

Surface brightness profiles for the M31 clusters were measured by combining
integrated photometry with star number counts (the
`hybrid' procedure described in Appendix~\ref{sec:artclust}). 
In the inner regions of the clusters, surface brightness profiles
were derived using the IRAF {\sc ellipse} package to fit circular
isophotes to the image data.
The isophote centers were fixed at a single value
for each cluster, with centers determined as the intensity-weighted centroid in a 
75 by 75 pixel box. 
Star counts were derived only from stars within specified regions of the CMD, with the
designated region varying by cluster depending on the age.
The details of the star counts for the GO-10818 clusters are
given by \citet{perina10};  for the GO-8296 clusters, star counts
were computed from background-subtracted CMDs 
\citep[Fig.\ 6 of][]{williams01} with positional data kindly
provided by B. Williams. The star counts were used for
radii $>7$~pc (40 pixels) from the cluster centers, and scaled to
linear intensity units ($L_\sun$~pc$^{-2}$) by matching the
counts and photometry over the overlap region 5--10~pc.
The same star counts were matched to integrated photometry
profiles in both red and blue filters, but with different scaling factors;
star count uncertainties were matched to the photometry
uncertainties by scaling as for the intensity.
No background subtraction was performed on the star counts.

\subsection{Profile-fitting methods}
\label{sec:fitting}

There are a number of possible choices for star cluster density profiles,
including \citet[hereafter King]{king66}, \citet[hereafter Wilson]{wilson75}, \citet{king62},
\citet[also known as `power-law' or `EFF']{elson87}, and \citet{sersic68}.
Unlike  the other three types of model profile, the
King and Wilson models have no analytic expressions for density or surface
brightness as a function of projected radius; profiles are
obtained by integrating phase-space distribution functions over all velocities 
and then along the line of sight, assuming spherical symmetry
\citep[for a review, see][]{mclaughlin03}. The King model is the most
commonly-used in studies of star clusters; however, \citet{mclaughlin05} showed
that, with data that extends to sufficiently large projected radii,
many Local Group clusters are better-fit by the more-extended Wilson models.
Globulars in NGC~5128 are also better-fit by Wilson models \citep{mclaughlin08a},
although an analysis using nearly identical techniques \citep{barmby07}
found that massive M31 globulars were better-fit by King models.
Taken together, these recent analyses showed that  fitting the \citet{king62},
\citet{elson87}, and \citet{sersic68} models did not add significant information
beyond that provided by the King and Wilson models, so we consider only
these two models in our analysis.

The King and Wilson models are single-stellar-mass, isotropic models
defined by phase-space distribution functions of stellar energy $E$:
\begin{equation}
f(E)\propto\left\{
\begin{array}{lll}
\exp[-E/\sigma_0^2] -1 &,\ E<0 &{\rm (King)}\\
\exp[-E/\sigma_0^2] -1 + E/\sigma_0^2 &,\ E<0 &{\rm (Wilson)}\\
0 &,\ E\ge 0\  &{\rm (both)}\\
\end{array}
\right.
\end{equation}
where $\sigma_0$ is the central velocity dispersion.
The effect of the extra term in the Wilson model $f(E)$ is to make
clusters more spatially extended. Both sets of models are characterized
by three parameters: a dimensionless central potential $W_0$, which
measures the degree of central concentration;
a scale radius $r_0$, which sets the physical scale; 
and a central intensity $I_0$, which sets the overall normalization.
For the King models, $W_0$ has a one-to-one correspondence with
the more-familiar concentration $c=\log(r_t/r_0)$, where $r_t$ is
the tidal radius at which the density $\rho(r_t)=0$. Possibly
contrary to intuitive expectations,  for two profiles with the same scale radius, 
the profile with a larger value of $c$ or $W_0$ declines more slowly.

Deriving the structural properties of the simulated clusters involved
fitting their projected surface density profiles to models using the
GRIDFIT program described by \citeauthor{mclaughlin05}
(\citeyear{mclaughlin05}, see also \citealt{mclaughlin08a}).
The program uses a grid 
of model density profiles, pre-computed for a range of values of $W_0$, then
finds the scale radius $r_0$ and central surface brightess $I_0$ to
minimize the weighted $\chi^2$ for each $W_0$; the best-fitting model is the one
with the global $\chi^2$ minimum. The model profiles are
convolved with the  instrumental PSF before comparison to the data.
Since no background subtraction was
performed on the star counts, the background level was 
determined as one of the parameters of the model fitting.
For a few clusters the fitting algorithm converged to unreasonably 
large or small values, and a fixed background
corresponding to the lowest level reached by the star counts
was subtracted before re-fitting; in general this procedure 
improved the reduced $\chi^2$ of the fits.

\subsection{Profile-fitting: results}
\label{sec:m31fits}

Figure~\ref{fig:m31profs} shows the cluster surface brightness profiles 
together with the best-fitting models. The parameters of the models are given
in Table~\ref{tab:fits}, corrected for extinction using the values of 
$E(B-V)$ given by \citet{perina10} or \citet{williams01}.
Conversion of filter-specific measurements
to the $V$-band is done using the transformations described in the appropriate
HST Instrument Handbooks; briefly, we compute the extinction-corrected color $(V-x)_0$, 
where $x$ is the observed-band magnitude, as a function of color in standard bands (e.g., $(V-I)_0$).
Ground-based integrated colors from \citet{rbc07} are used for 
the standard-band colors, to avoid iteration; uncertainties of 0.1~mag in $(V-x)_0$ are assumed
and propagated through the parameter estimates.
As previously shown by \citet{mclaughlin05},
differences between Wilson and King model profiles occur primarily in the 
outer parts of  cluster profiles, where our signal-to-noise is low. 
The similarity between model profiles also means that, in general, the best-fit models 
of the two families have very similar $\chi^2$, with
no strong systematic preference for one model or the other. 
Typical $\chi^2$ values are 85--90;
with $\sim 20$ datapoints and 3 or 4 model degrees of freedom, the
resulting reduced  values are $\chi^2_{\nu} \sim 6$.
This indicates that the uncertainties produced by integrated photometry are 
likely underestimates, and one reason may be that these uncertainties do not 
account for the uncertainty in the background level.
Rather than modify the uncertainties to achieve 
$\chi^2_{\nu} \sim 1$, we modified our use of $\chi^2$ in computing 
parameter uncertainties \citep[see also][]{mclaughlin08a}.
We scaled the reduced $\chi^2$ values such that the best-fit model had $\chi^2_\nu \equiv 1$.
The 68\% confidence limits on the parameters are then the minimum and maximum
values found in the set of models with $\chi^2_\nu \leq 2$.
This rescaling gives more realistic estimates of the parameter uncertainties
than would otherwise be the case.

How robust are the physical parameters derived from our model fits?
One way to estimate this is to compare various fits to
the same cluster. Although $W_0$ and $r_0$ have slightly different
meanings in King and Wilson models and cannot be directly compared,
some derived quantities such as  the half-light radius and total luminosity 
are directly comparable. 
For all clusters we have profile data in two 
different bandpasses, although the outer parts of the profile, derived
from number counts, are the same in both.
There are physical reasons why profiles might change with
wavelength (e.g., mass segregation, differential reddening), but comparison
of model fits in different filters is a useful sanity check.
Figure~\ref{fig:model_comp} shows this comparison: the scatter between
filters is 0.2--0.3~dex. A similar comparison between fits for M31 globular clusters
by \citet{barmby07} found a much smaller scatter, probably because that
work analyzed bright clusters, using much deeper data.
Figure~\ref{fig:model_comp} also compares $R_h$ and $L_V$ between Wilson and
King models. The scatter is again rather large, 0.15--0.25~dex,
with the Wilson models offset to larger values. To some extent this is to be
expected, since Wilson models have larger halos; however  some
of the Wilson model values (e.g., $R_h>50$~pc for B015D, B257D, B321,
B376, and B448) are physically implausible, because the model-fitting
resulted in a very large values of the central potential $W_0$.
We do not completely understand the reason for this but speculate
that it may be related to the 
combination of the additional power in the haloes of Wilson models
and the low signal-to-noise of the profiles in the same region.
These results indicate the limitations of our relatively shallow data, and the 
limited precision of the model measurements will need to be kept in mind 
during the following analysis. 

For the analysis in the remainder of this paper,  we use only a single set of
model parameters per cluster. Because the King models have fewer 
implausible values, and also somewhat less scatter between filters,
we use on the King model parameters for the present cluster sample.
Our results in Appendix~\ref{sec:artclust} indicate that King model fits may be more
robust than Wilson model fits in the case where background levels are uncertain,
even where the underlying cluster profile is actually a Wilson model.
Using King models also allows us to compare the present sample to the
combined sample of M31 globulars analysed in \citet{barmby02,barmby07}:
all of that sample has King fits while only about one third has Wilson-model fits.
Because the focus of this paper is the young M31 clusters, dominated by blue stars,
we use the F439W or F450W-band measurements in preference to those from 
the redder filters.

The left panel of Figure~\ref{fig:lum_age_struct} shows the properties of the present sample
of clusters as a function of luminosity. Four clusters (vdB0, B327, B342, B368)
stand out as having very  high central surface brightnesses; all except
B327 also have correspondingly high concentrations. Figure~\ref{fig:m31profs}
shows that the cores of these clusters do not appear to be resolved in our data. This
could be due to the short exposure times: if the central cluster light is dominated by a
few bright stars, the true integrated profile could be very difficult to recover.
Structural parameters for these clusters are uncertain.
Figure~\ref{fig:m31profs} also shows that the three M31 young clusters with the 
largest inferred half-light radii (B015D, B321, B448) have relatively low
contrast against the resolved stellar background of M31, so it is possible
that the number counts include some field stars and the resulting $R_h$ values are overestimates. 
The old cluster NB16 has a much smaller $R_h$ and total luminosity 
than the other members of the sample: this cluster is projected on
the M31 bulge and its outer stars may be lost against the bright background.
These issues highlight the limitations of our dataset for the kind of structural
analysis we are attempting, but the generally good match of model profiles
with the observational ones gives us confidence that the cluster parameters
we measure are reasonable.

Analyzing the physical properties of M31 young clusters requires
converting the observed flux-based measurements to luminosities
and mass-linked quantities. 
Conversion from luminosity to mass is done using $V$-band
mass-to-light ratios from the population synthesis models of \citet{bruzual03},
assuming a \citet{chabrier03} IMF and solar metallicity for all but
the oldest clusters. Table~\ref{tab:hstdat} lists the assumed ages for all clusters:
those given by \citet{perina10} for the young clusters from GO-10818, by \citet{williams01} 
for the clusters from GO-8296, and assumed ages of 13~Gyr
for the clusters B083, B222 and B347, B374 and NB16.
We assume uncertainties of 10\% in $M/L_V$ and 
propagate these through the parameter estimates.
While using $M/L_V$ ratios determined directly from 
measured velocity dispersions would avoid the reliance on models, 
velocity dispersions are  not available for most of the M31 clusters
considered here. The use of a single set of population synthesis 
models  also facilitates comparison of clusters in different galaxies; the comparison
data for other galaxies, \citep{mclaughlin05, barmby07, mclaughlin08a}
also used the same model mass-to-light ratios. Tables~\ref{tab:mod_param} 
and \ref{tab:mod_dynam} give various derived  parameters for the best-fitting
models for each cluster \citep[the details of their calculation are given by][]{mclaughlin08a}.
Recently, \citet{kruijssen08b} have discussed of star cluster mass-to-light ratios
due to preferential loss of low-mass stars with cluster age.  This effect is expected to be most
important for old clusters, and we have used the \citeauthor{kruijssen08b} models to
confirm that the change in $M/L$ for young clusters is minimal ($\lesssim 20$\%).
Since our focus in this paper is the young M31 clusters,
we therefore do not correct for this effect.

\section{Discussion: young and old clusters in M31 and other galaxies}

Using star clusters as markers of the history of galaxies is aided by
knowing how the clusters' structural properties change with age
and environment. 
Although absolute ages of star clusters are notoriously difficult to determine,
relative ages are more straightforward, and all of the clusters in our sample
have ages estimated by CMD fitting \citep{williams01,perina10}. Can we see evidence
for changes in cluster properties with age? In the right panel of 
Figure~\ref{fig:lum_age_struct},
structural properties for the M31 young clusters are shown as a function of estimated age.
None of the properties plotted depends on mass-to-light ratio, which is 
strongly dependent on age. Although our sample is small and covers a limited 
range in age, there is an interesting
hint that central surface brightness becomes fainter and concentration decreases
as age increases. This is consistent with the increase in core radius with age 
for MC clusters noted by \citet{mackey03}.
Figure~\ref{fig:mu0_c_age} explores this further by plotting
$\mu_0$, $c$, $R_c$, and central mass density $\rho_0$ 
for both  the M31 young clusters  and young clusters
in the Magellanic Clouds. While the MC clusters also show a trend for 
central surface brightness to fade with age, it is much weaker than 
the trend implied by the M31 clusters alone, and the high-surface-brightness
M31 clusters appear to be outliers (possibly artifacts due to the limited spatial
resolution). Since the central mass density shows very little trend with age,
the central surface brightness trend is likely due to fading of stellar population
and the (weak) increase of core radius with age. The dashed line in the
central surface brightness panel shows the effects of
mass-to-light ratio change predicted by the \citet{bruzual03} models with a \citet{chabrier03} 
IMF and solar metallicity; the slope shows a reasonable match to the cluster trend.

Figure~\ref{fig:mu0_c_age} shows that, with a few exceptions, the young M31 
clusters have similar spatial structure to young clusters in the Magellanic Clouds. 
A number of young massive clusters have recently been identified in the
Milky Way; \citet{pfalzner09} compiled size and mass measurements of
these clusters \citep{figer08,wolff07} to argue that cluster evolution occurs along 
two well-defined tracks in the density-radius plane.
Using the conversion between Milky Way cluster size measurements 
and half-light radii described in Appendix~\ref{sec:artclust}, we have compared cluster
half-light radii  and ages for the young Milky Way clusters together with the M31 and MC
clusters in Figure~\ref{fig:age_size}.  The M31 and MC clusters have similar
sizes to the `leaky' Milky Way clusters but lie 
on the extrapolation of the age-$R_h$ trend of the `starburst' MW clusters.
This suggests that the starburst clusters (which tend to be more massive)
are perhaps closer to being analogs of the young massive clusters in other galaxies.
We speculate that  the two evolutionary paths of \citet{pfalzner09} may be 
simply due to extinction effects, with the `starburst'  clusters having left their host cocoon 
and the `leaky' clusters still affected by excessive extinction in their outer regions 
(projection effects may also be important). This would imply that starburst clusters
are more easily identified in external galaxies,  explaining the reasonable match
between extragalactic young clusters and Milky Way starburst clusters.

An important question in the study of young massive clusters is 
whether they will eventually become old massive clusters resembling
the globular clusters we see today in the Galaxy. Once formed,
star clusters have no easy way to gain mass, but they do have
a number of ways to lose mass or even be completely disrupted
\citep{spitzer87,vesperini98,lamers06}. We have computed dissolution times
for our cluster sample considering the effects of both the stellar and dynamical 
evolution of star clusters through time. These calculations explicitly account for 
age, metallicity, and half-light radius of all sample star clusters, and treat the 
effects of evaporation of low-mass stars, mass loss due to stellar evolution, 
encounters with spiral arms and giant molecular clouds following in part 
the prescriptions of \citet{lamers05}  and \citet{lamers06}
The results are shown in Figure~\ref{fig:t_dissolve}.
All clusters have dissolution time greater than their ages;
however, for 2 young clusters (B321, B342) and the old
cluster B374 these quantities are nearly equal, 
suggesting that they are in the process of dissolving.
On average, the young clusters' dissolution times are too short to
expect them to become old ($>10^{10}$~yr) clusters. However, 
a few have $t_d>1$~Gyr and, if they
avoid collisions with giant molecular clouds,
might survive to become sparse old globulars.
In general, the dissolution times confirm the importance of cluster dissolution 
to the evolution of the star cluster mass function
\citep[see also, e.g.,][]{gnedin97a,gieles09}.
Lower-mass and/or more-diffuse clusters in M31, such as those discovered
by \citet{krienke07,krienke08} and \citet{hodge09}, would be even more likely
to dissolve.

Work to date suggests that the structural parameters of old star clusters in several nearby galaxies
show only a weak dependence on environment \citep{barmby07},
and the comparisons above indicate that young clusters in different galaxies
are also similar. How do young and old clusters compare?
Figure~\ref{fig:struct_comp} shows cluster properties as a function of
mass for  M31 young clusters, Magellanic Cloud young clusters and Milky Way 
globulars \citep{mclaughlin05}, M31 globulars \citep{barmby02,barmby07},
and recently-discovered extended M31 halo clusters \citep{huxor05}.%
\footnote{Mass measurements for all clusters are derived using mass-to-light ratios.
As discussed in \S\ref{sec:m31fits}, these ratios are affected by cluster dynamical
evolution. Correcting for this effect is non-trivial and beyond the scope of this paper; 
however the results of \citet{kruijssen08a} imply that doing so would increase the
spread of the old clusters' mass distribution and shift it to lower masses.}
The joint mass-age distribution of the clusters differs by galaxy:
some of this is due to complex selection effects (e.g., the M31 globular sample is incomplete 
and biased toward more luminous clusters, and the sample of Milky Way YMCs is also
incomplete), but there are hints of real differences between galaxies; see
\citet{perina10} for a more detailed discussion.
The properties of the five old clusters in our sample are similar to those
of M31 and Milky Way globulars, while
the properties of  M31 young clusters overlap with those of both
the young Magellanic Cloud clusters and the low-mass
Milky Way globular clusters. Thus the M31 young clusters do not appear to
be fundamentally different types of object from those already known.
On average, the younger clusters have larger sizes and higher concentrations 
(where larger $c$ implies a larger tidal radius for the same scale radius)
than old clusters of the same mass.
The young clusters therefore have larger tidal radii, which makes
them more susceptible to dynamical destruction: small-$r_t$
clusters are more likely to survive to old age.
The larger spread in properties of low-mass clusters
compared to higher-mass clusters may indicate lower data quality for these
fainter objects, rather than an intrinsic difference in properties. 

By now it is well-known that old star clusters in the Milky Way and other
galaxies describe
a `fundamental plane' (FP) in structural properties 
\citep{djorgovski95,djorgovski97}, although the
separation of clusters from other types of objects has become less
well-defined in recent years. The results of \citet{bastian06}
and \citet{kissler-patig06} indicate that young massive clusters
fall on similar fundamental planes to those of old clusters. 
Those results make use of cluster velocity dispersions, while
in this work, we must use mass-to-light ratios
from population synthesis models applied
to the photometry instead of independent mass estimates.
The upper-right panel of Figure~\ref{fig:struct_comp} shows one
view of the FP, as defined by \citet{mclaughlin00}. 
The old clusters in our sample fall nicely on this
relation, as do most of the younger clusters.
The observed correlation between mass and binding energy $E_b$ 
is expected, since by definition $E_b= f(c) M^2/R_h$ where $f(c)$
is a weak function of cluster concentration $c$. 
However, the tightness of the correlation shows that there is
very little relation between young cluster mass and $R_h$ (see also lower-right panel), 
and no offsets in the basic properties of the cluster shapes between old and young clusters.
 
Figure~\ref{fig:cluster_fp} shows a different view of the fundamental plane,
more akin to the parameters usually shown for elliptical galaxies
\citep[see also][]{mclaughlin03,strader09}. The left two panels show the 
surface-brightness-based fundamental plane relations, with
a large offset between the young M31 and MC clusters (light grey symbols)
and the old clusters. This is to be expected because of the young clusters'
lower mass-to-light ratios. When we instead plot quantities related to the 
mass density (right panels), the young clusters fall on the same relations
as the old clusters. The tightness of the relations primarily reflects the use of
mass-to-light ratios to compute both central velocity dispersion $\sigma_0$
and mass density $\Sigma$. Again, however, the lack of offset and similar scatter
between the young and old clusters confirms their similar overall structures.
Recent measurements of M31 globular clusters' mass-to-light ratios \citep{strader09}
have shown that these clusters do follow the FP relations as expected
from model mass-to-light ratios. Similar measurements for young clusters
should show whether  young clusters do the same.
If so, this would indicate that the FP
reflects conditions of cluster formation and is not merely the end product of
cluster dynamical evolution.

\citet{bonatto05} argue that Milky Way open clusters fall on a plane in 
the three-dimensional space of total mass, core radius, and 
projected core mass density. We can compare this space to the FP using
with an approximate relation between mass and central velocity dispersion.
The least-squares fit for the young MC clusters (the most populous sample of
young clusters available) gives 
$\log \sigma_0 = 0.34 \log M -1.38$; combined with the \citet{bonatto05}
cluster parameters, we find that the Milky Way open clusters fall approximately on the
other young clusters with $\Sigma_0\sim 10^2$~M$_\sun$~pc$^{-2}$  
in the top right panel of Figure~\ref{fig:cluster_fp}.
This suggests that the Milky Way open cluster plane indicated by
\cite{bonatto05} may in fact be the same FP defined by other star clusters,
which have projected mass densities higher by up to four orders of magnitude.
As \citet{bonatto05} discuss, this result remains to be confirmed with
large samples, but it is certainly intriguing in its implications for a 
`universal' star cluster fundamental plane.

\section{Summary and Directions for Future Work}

This series of papers has established that a sample of candidate young
star clusters in M31 are indeed young, massive clusters, with properties similar
to those of other young clusters in Local Group galaxies. Our current data
does not allow us to detect the extended haloes characterized by Wilson models
and seen in other young clusters; the more compact King models provide adequate
fits to the data. The structural parameters measured in this paper show the
M31 clusters to be typical young clusters, with masses of $10^{4-5}$~M$_\sun$,
half-light radii of 3--20~pc, and dissolution times of $<5$~Gyr.  While the
basic similarity between young clusters in different Local Group galaxies,
and between young and old clusters, seem well-established, many questions remain.
What is the precise form of the age-size relation? Do cluster mass-to-light ratios
evolve with age as predicted by dynamical and stellar evolution models? What
fraction of the stellar disk in galaxies is comprised of dissolving clusters? Is there 
a relation between the cluster formation and local star formation rate, or other galaxy
properties? Large cluster samples with high-quality data will be needed to address 
these and other questions about the relationship and history of star clusters and
their parent galaxies.

Facilities: \facility{HST (WFPC2)}

\acknowledgments
We thank D.E. McLaughlin for the use of the GRIDFIT software.
P.B. acknowledges research support through a Discovery Grant from the Natural Sciences
and Engineering Research Council of Canada and an Ontario Early
Researcher Award.
S.P. and M.B. acknowledge the financial support of INAF through 
PRIN 2007 grant no.\ CRA 1.06.10.04 ``The local route to galaxy
formation.'' J.G.C. acknowledges support from NASA from grant GO-10818.
T.H.P acknowledges support through the Plaskett Research Fellowship at the Herzberg Institute of Astrophysics.

\appendix
\section{Artificial cluster tests}
\label{sec:artclust}

Deriving surface brightness profiles of star clusters in Local Group 
galaxies requires careful analysis. The clusters are only partially
resolved into individual stars, and they are observed together with
a galactic background which may also be resolved into stars.
The purpose of this section is to investigate the best methods for 
extracting structural parameters of `semi-resolved' clusters, particularly
from relatively shallow images, and to quantify the uncertainties
of those parameters. This can best be done by analyzing profiles derived from 
images of artificial clusters whose structural parameters are known.
A related study by \citet{noyola06} simulated integrated photometry from HST observations
of  Galactic GCs; however the focus of that study was on
recovering the structure of cluster cores rather than overall structure.
\citet{bonatto08} also carried out a similar study, but considering only \citet{king62} models
for Galactic clusters.

The first step in analyzing simulated star cluster profiles is to determine
the type of model profile and range of parameter space to be covered.
The analysis of \citet{mclaughlin05} showed that Wilson models were 
adequate to describe both Milky Way and Magellanic Cloud cluster profiles, so we 
chose this set of models for our artificial clusters. 
Since we are interested in differences between young and old clusters
we examined the distribution of  scale radius $r_0$ and
central potential $W_0$ for both young and old Magellanic Cloud clusters as given by \citet{mclaughlin05}:
$W_0$ ranged from 1 to 10  with a typical value $W_0\approx 5$
while $r_0$ ranged from 0.2 to 20~pc with a typical value  $r_0\approx 2$~pc.
The range of implied half-light radii is 1--35~pc.

Our artificial clusters were generated from Wilson profiles with   
8 values of $r_0$ between 0.5 and 11~pc, and  9 values of $W_0$
between 2 and 10.  For each $(W_0,r_0)$ pair
we generated clusters with four different population sizes: $N_*=100, 300,1000, 3000$.
The stars' projected spatial positions  were generated by selecting the
projected radial coordinate from the probability distribution associated
with the Wilson profile
\begin{equation}
p(R) = \frac{R\, \Sigma_{W_0,r_0}(R)}{\int_0^{R_{\rm max}} \Sigma(R^\prime)R^\prime dR^\prime}
\end{equation}
and generating the angular coordinate $\theta$ at random.
The stars' luminosities were generated by selecting from an observed `young cluster'
luminosity distribution, uncorrected for completeness. 
The distribution was generated by combining the observed magnitudes of stars in the
four most populous clusters in the GO-10818 program
(VdB0, B257D, B475, B327). Separate luminosity distributions were
used in each of the two observational bands.

The specific observations being modeled are the same as those
in the GO-10818 program. We generated images of the simulated clusters
by inserting artificial stars modeled with the appropriate PSF near 
the center of a WFPC2/PC image of a field in M31. 
The background images used were the observations of `B195D' 
from the  GO-10818 program;  the PC chip was essentially empty in this 
observation because of an error in the input coordinates
\citep[for details, see][]{perina09}. This field is located
in the south-west disk of M31.
Figure~\ref{fig:pix_f450} shows a sample of the simulated 
cluster images, together with some sample M31 clusters for comparison. 
The simulated clusters cover a wider range of properties
than the real clusters: some of the simulated clusters were in
fact not visually apparent  in the images.  These `clusters' had few stars
($N_*=100$ or $N_*=300$) and very large half-light radii, more akin to
dwarf galaxies than to objects recognizable as star clusters. They are
not considered further in this analysis.

Surface density profiles for the simulated clusters were derived in several
different ways. The first method (`number counts'),
derived the surface density as simply the number of stars per unit area
in annular bins.  Since the locations of all
stars are known precisely for the simulated clusters, this method represents
the best possible data for surface density profiles. Deriving structural parameters
from such data tests the fitting routine itself and also the extent to which 
density profiles can be derived from a limited number of stars.
Stars were counted in overlapping
annular bins of width 3 pixels (0.5~pc) inside a radius of 20 pixels (3.4~pc)
and width 10 pixels (1.7~pc) outside this radius. 

For real star clusters, crowding limits the ability to resolve individual
stars and hence derive surface density profiles through number counts.
We also derived surface density profiles of clusters using
isophotal photometry with the IRAF {\sc ellipse} package,
similar to the method described in \citet{barmby07}. 
We refer to this  as the `integrated photometry'
method.  We also combined the number count and integrated photometry methods
in a `hybrid' method similar to that used by \citet{federici07}.  This involves
matching  the intensity scales of the two profiles by fitting both profiles to smooth curves
in the region $r=5-10$~pc. 
The switch-over from integrated photometry to number counts was made at a 
radius of 7~pc (40.6 pixels), where in general both types of profile had good signal to noise.

Wilson models were fit to the artificial cluster data using the GRIDFIT
program described in \S\ref{sec:fitting}. 
As for the real clusters, instrumental PSF profiles were convolved
with the model profiles before comparison to the data. Unlike the real
clusters, however, the background level for the artificial clusters was fixed
at zero. For clusters of all sizes, the number count input
returned fitted parameters in good agreement with the input parameters.
The offsets between input and output parameters
are (mean $\pm$ standard error) $\Delta W_0 = (W_{\rm 0,in} -W_{\rm 0,out})/W_{\rm 0,in} = 0.06\pm 0.02$ and
$\Delta r_0 = (r_{\rm 0,in} -r_{\rm 0,out})/r_{\rm 0,in} =-0.13\pm0.03$~pc.
As expected, the larger-$N_*$ clusters return more accurate values,
with scatter 2--3 times lower for $N_*=3000$ than for $N_*=300$ clusters.
Figure~\ref{fig:comp_artphot} compares the best-fit and input structural 
parameters of the simulated clusters for the integrated photometry 
and hybrid methods. Particularly for clusters with larger input $r_0$, integrated
photometry alone tends to result in
overly-large values of $W_0$ and overly-small values of $r_0$.
For these clusters, the distinction between profiles of different $W_0$ 
occurs at a point in the radial profile where the density of stars is too low for the {\sc ellipse}
algorithm to converge. The addition of number count data beyond this
point improves the fit, as the figure shows.
For integrated photometry alone, $\Delta W_0 =-0.56\pm0.07$ and $\Delta r_0 =0.24\pm0.04 $~pc; 
for the hybrid method, $\Delta W_0 =-0.02\pm0.02$ and $\Delta r_0 = -0.05 \pm 0.03$~pc.

When fitting model profiles to cluster data, the correct model family is not 
not known {\it a priori}. What happens if artificial `Wilson' clusters are fit with
King models instead? We tried this experiment with our artificial clusters
and were surprised to find that, except for a handful of objects, the two model
families returned nearly identical $\chi^2$ values: the median fractional
difference $(\chi^2_{\rm K} - \chi^2_{\rm W})/\chi^2_{\rm W}  = 0.01$.  While
the meaning of model parameters such as the scale radius $r_0$ 
differs between model families, some derived quantities such as the core and
half-light radii ($R_c, R_h$: see Table~\ref{tab:mod_param} for description)
are directly comparable. Figure~\ref{fig:art_king_wils} shows this comparison.
There is very good agreement between the two model families in
measurements of core radii, and reasonable agreement in measurements of half-light radii.  
The agreement in $R_h$ is poorer for the largest clusters ($R_h\gtrsim 20$~pc, 
a larger size than usually seen in real clusters),  where the
King models return smaller sizes than the Wilson models.
This is consistent with the results of \citet{mclaughlin05} who found that 
the two model families gave generally consistent results for Milky Way 
and Magellanic Cloud clusters as long as the
radius of the last data point $R_{\rm last} \gtrsim 5R_h$.

The situation of observational profiles with a limited radial range bears further investigation. 
The analysis of simulated
clusters to this point has not considered the effects of background level fluctuations.
The GRIDFIT code is able to fit a constant background level added to the intensity profile,
and we verified through simple experiments 
that input values were correctly recovered. However, the limitations of short
exposures and small-number statistics suggest that determining the correct
background level---and thus being able to correctly trace cluster profiles out
to large projected radii---will be much more difficult for the real cluster data.
We therefore experimented with removing points in the profile data beyond
$R_{\rm last}=1$, 2, and $5R_h$ (where $R_h$ was computed from the input model profile)
and fitting both King and Wilson models to the remaining points. 
As expected, recovery of the input cluster parameters
was better for the more extensive profiles, for both model families. 
For $R_{\rm last}=1$, both model families returned $R_h$ values that
were, on average, larger than the input.
Some model fits were `catastrophic failures', with $R_h({\rm out}) > 2R_h({\rm in}) $; 
this situation usually occurred for  clusters where the number of profile data points 
was $<10$. Interestingly, for all three values of $R_{\rm last}$, 
King model fits had fewer catastrophic
failures than Wilson models, and also slightly smaller 
scatter in the difference between fit and true parameters.
Since the primary difference between King and Wilson 
model profiles is the more extended halo of the latter, 
this suggests that King models may be a better choice for
fitting noisy cluster profiles.

Finally, we considered the issue of comparison between different measurements
of star cluster size. While Milky Way globulars and extragalactic clusters are
most often characterized with half light or core radii, recent complilations of data
for massive young Milky Way clusters \citep{figer08,wolff07} measure cluster size as the
mean or median distance  ($\langle R \rangle$ or $\tilde{R}$) of the cluster stars from the 
geometric centroid.
Since these young Milky Way clusters may well not be dynamically relaxed \citep{goodwin06},
it may not make sense to fit the same types of dynamical models to them as to old
clusters, but it is still desirable to find a way to compare sizes between groups of
clusters. Since we know the positions of all stars in our artificial clusters, we can easily
compute the statistical measurements of size  for our
model clusters, and compare them to (model values of) $R_c$ and $R_h$. 
$\langle R \rangle$ and $\tilde{R}$
are very well-correlated for all of our model clusters, with a best-fit linear relation
$\tilde{R}= 0.67\langle R \rangle-0.36$. The correlation between 
$\langle R \rangle$ and $R_c$ is rather poor (unsurprising as $R_c$ depends critically
on the exact shape of the cluster profile), but there is a good match between
$\langle R \rangle$ and $R_h$ for models which are not too extended ($W_0\lesssim 6$).
Figure~\ref{fig:mean_rh} shows the data and least-squares fits:
$\langle R \rangle= 0.77R_h +0.23$, and $\tilde{R} = 0.53R_h +0.10$.
We conclude that, with some scaling, the 
mean or median projected separation of stars from a cluster center
are reasonable proxies for the half-light radius.

\bibliographystyle{aas}

\clearpage

\begin{figure}
\plotone{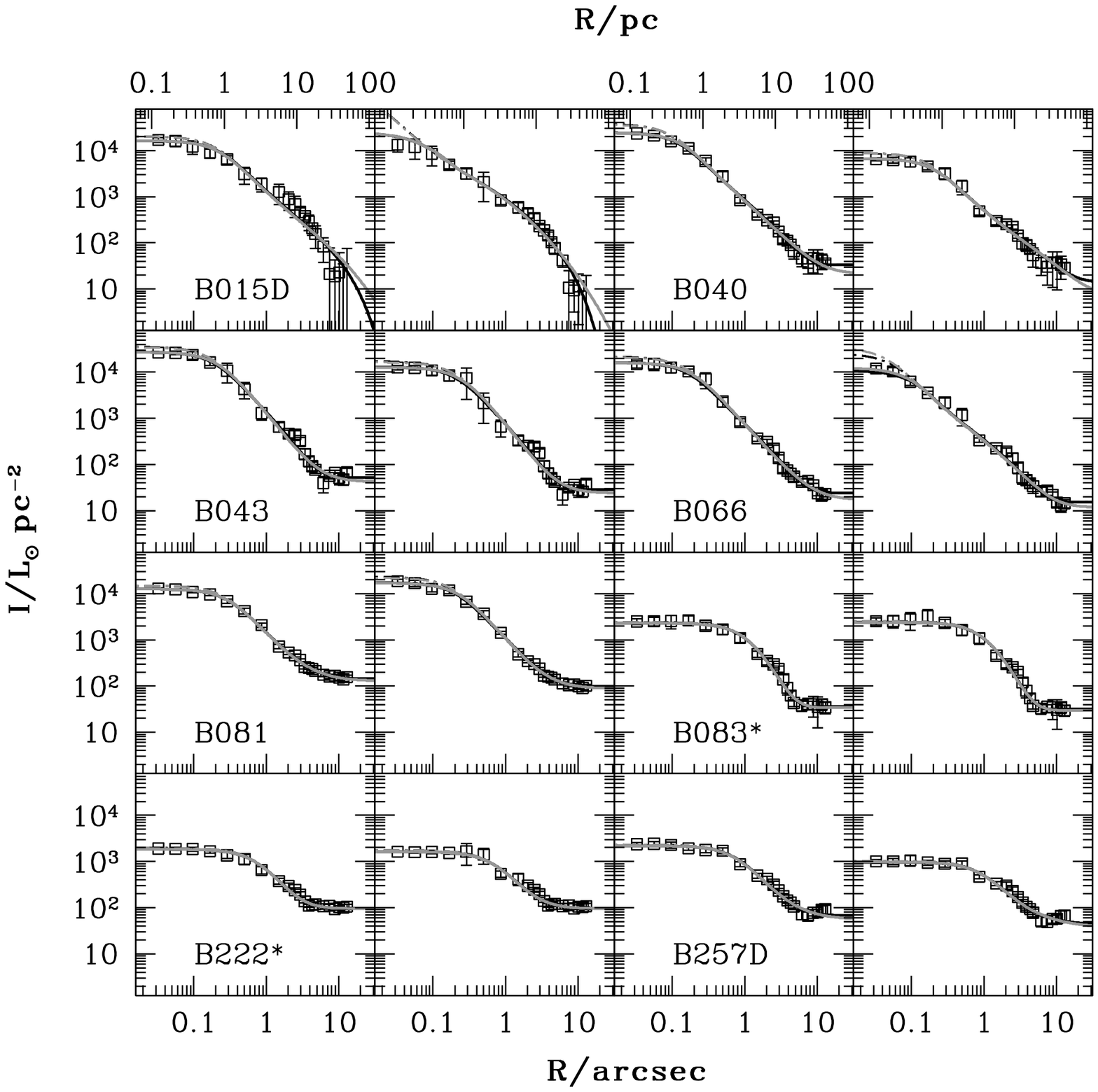}
\caption{
M31 cluster surface brightness
profiles together with the best-fitting models.
Each cluster is shown in two sub-panels, with the
bluer filter (F439W or F450W) on the left 
and the redder filter (F555W or F814W) on the right.
Clusters with an asterisk after their names are
likely to be old.
Black lines are best-fitting \citet{king66} models;
grey lines (most are directly over the black lines)
are best-fitting \citet{wilson75} models.
Solid lines are model profiles after convolution
with the PSF; dash-dot lines are profiles before
convolution.
\label{fig:m31profs}}
\end{figure}

\setcounter{figure}{0}
\begin{figure}
\plotone{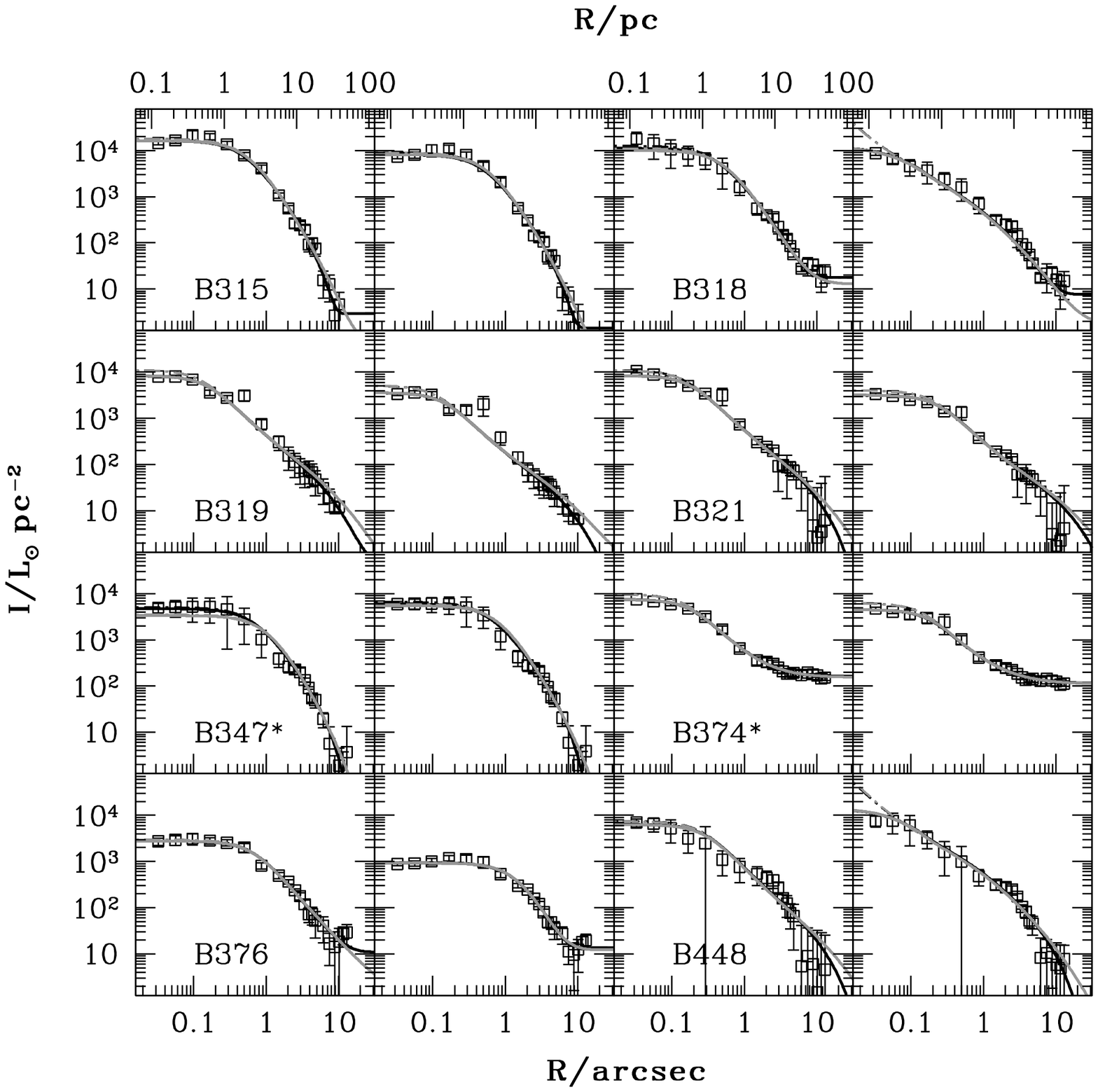}
\caption{Continued}
\end{figure}

\setcounter{figure}{0}
\begin{figure}
\plotone{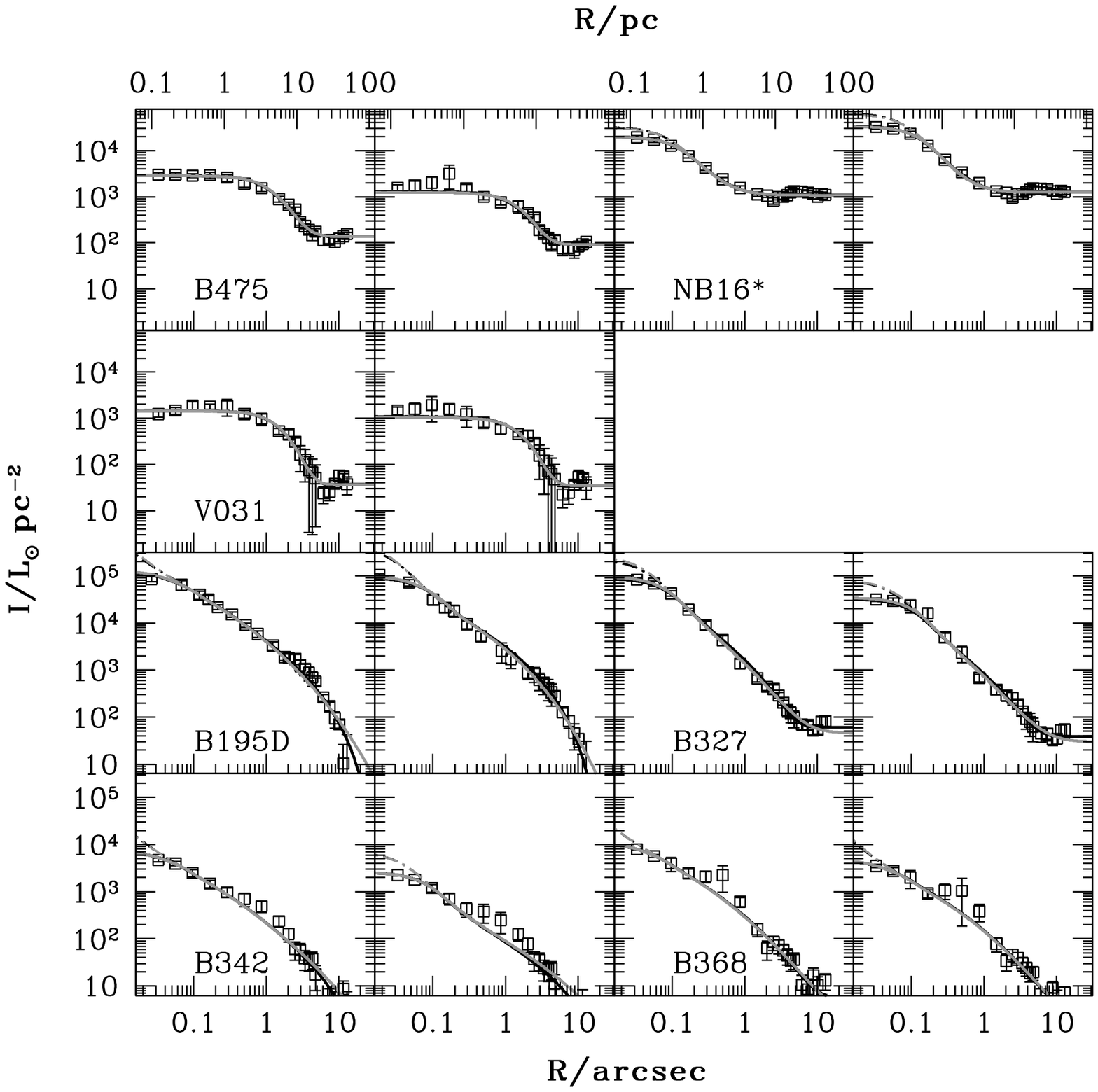}
\caption{Continued. Note that the last four clusters are plotted with a different vertical scale.}
\end{figure}

\begin{figure}
\plotone{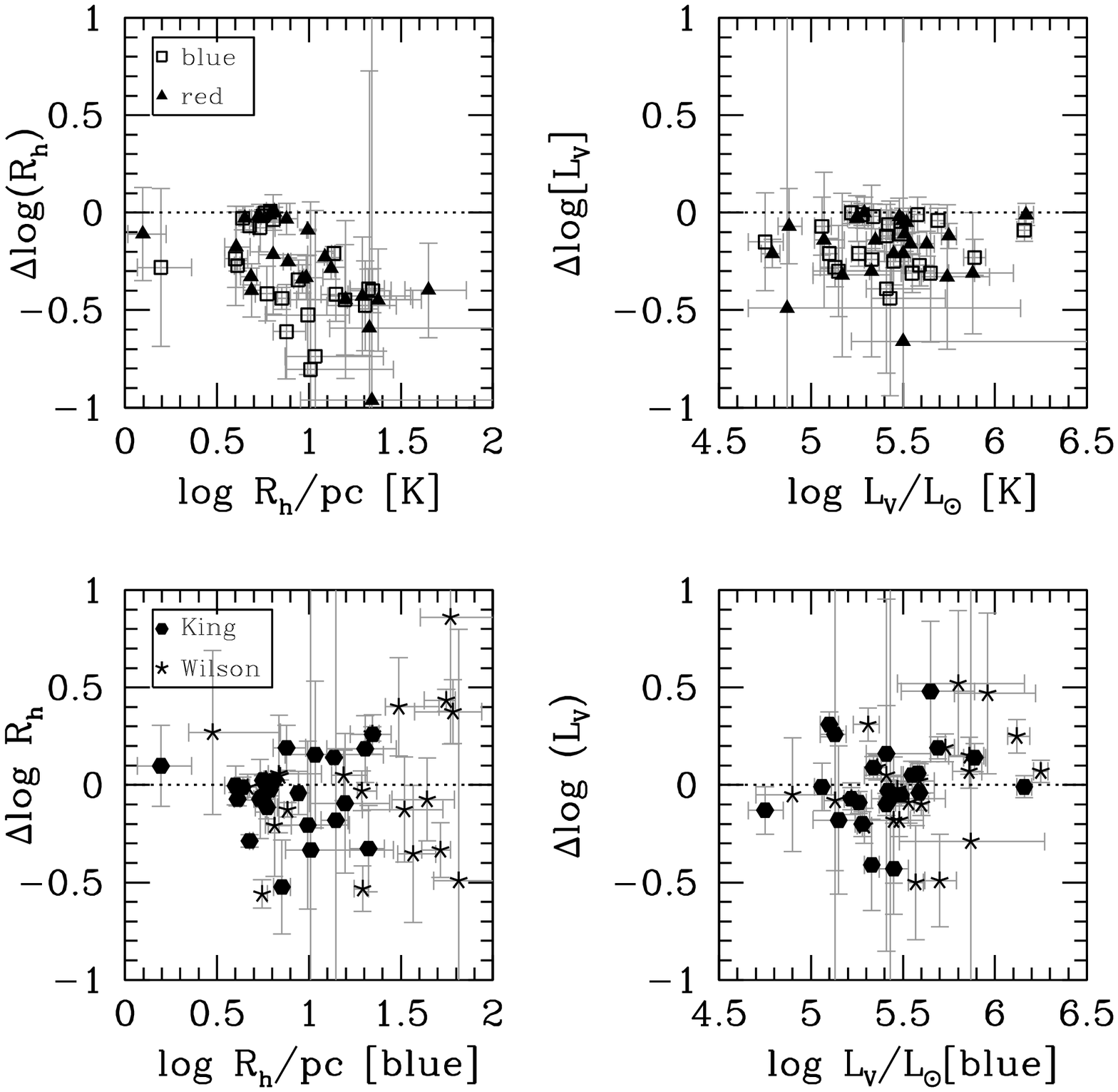}
\caption{Comparison of half-light radii and total luminosity (converted
to the $V$ band) for Wilson
and King models fit to surface brightness profiles of M31 young clusers.
Bottom: comparison between observations of the same cluster in
different filters (hexagons: King models, stars: Wilson models).
Top: Comparison of Wilson and King model fits to the same cluster
(squares: red filter, triangles: blue filter).
\label{fig:model_comp}}
\end{figure}

\begin{figure}
\plotone{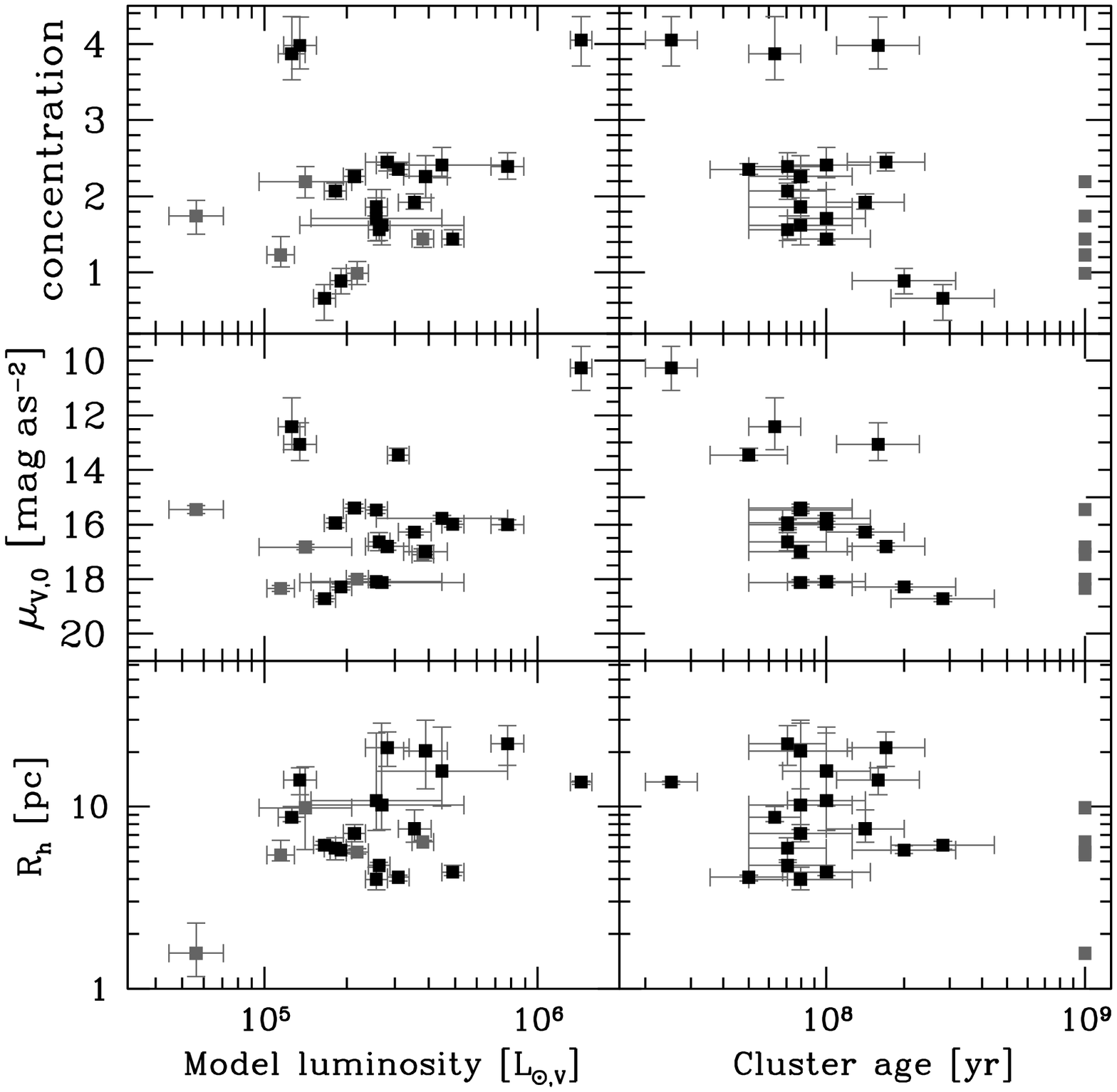}
\caption{
Concentration index, central surface brightness, and half-light radius  
for M31 young clusters as functions of total model luminosity (left)
and estimated age (right). The old clusters are shown with gray symbols;
although their ages are estimated at $>10^{10}$~yr, they are plotted 
at $10^9$~yr in the right panel to condense the horizontal axis scale.
\label{fig:lum_age_struct}}
\end{figure}

\begin{figure}
\plotone{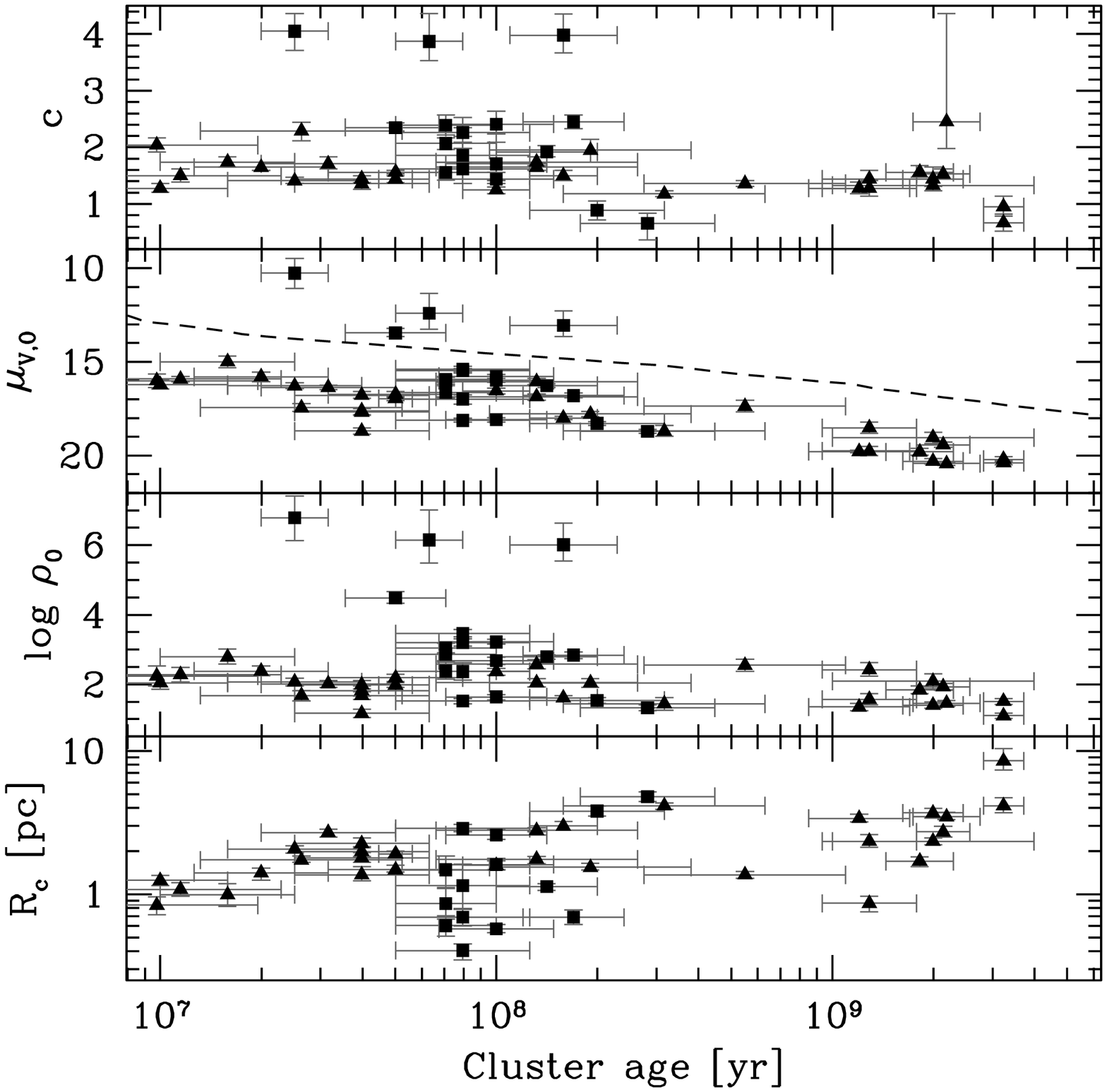}
\caption{
Concentration index, central surface brightness, and central mass
density for M31 (squares) and Magellanic Cloud (triangles)
young clusters as functions of estimated age.
The dashed line in the central surface brightness panel shows
the expected change in surface brightness due to changes in
mass-to-light ratio with age (vertical normalization is arbitrary).
\label{fig:mu0_c_age}}
\end{figure}

\begin{figure}
\plotone{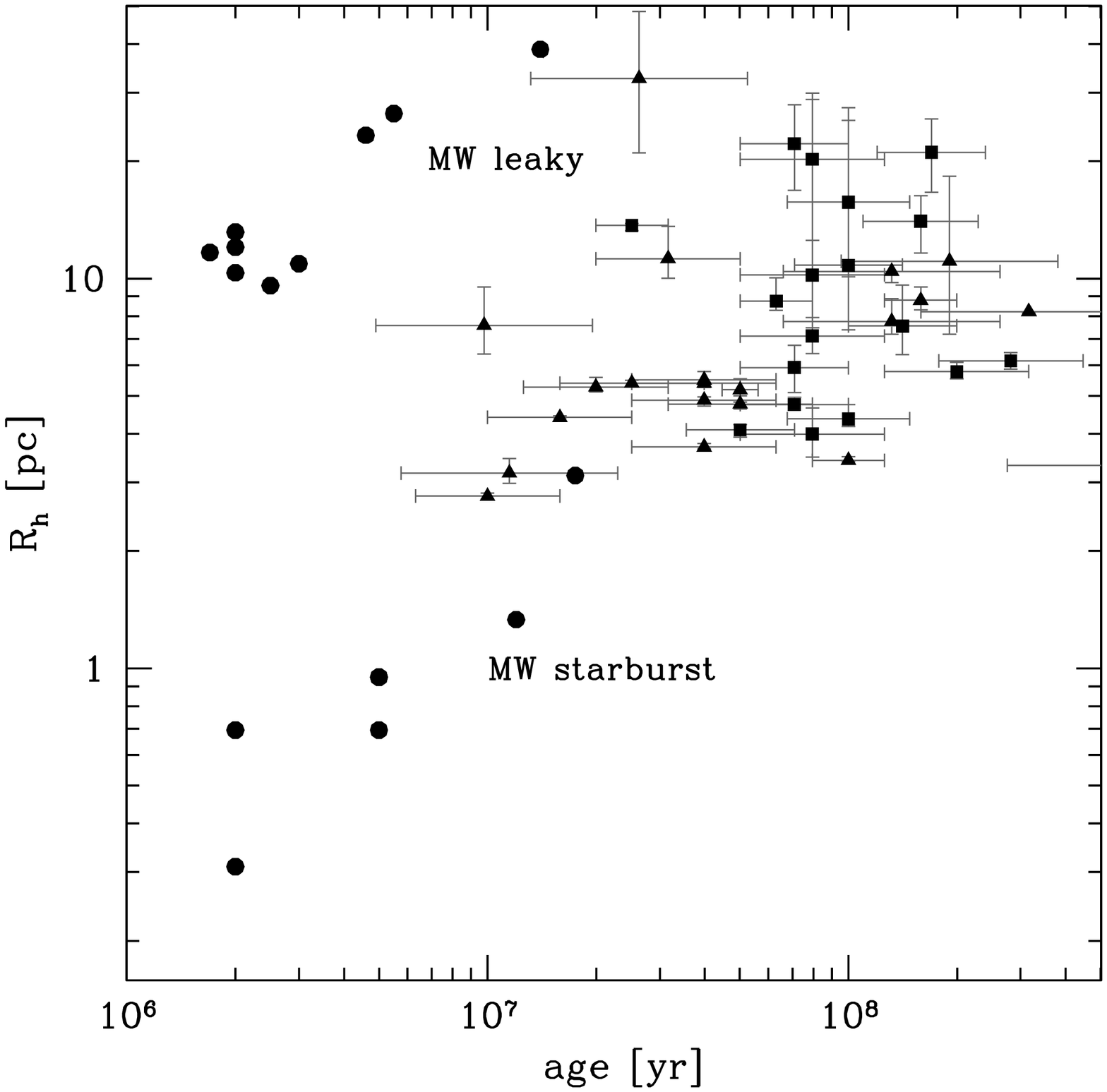}
\caption{
Young star cluster ages and sizes.
Squares:  M31 clusters from the present sample; 
triangles: young Magellanic Cloud clusters; 
circles: young massive Milky Way clusters from  \citet{figer08}
and \citet{wolff07}. The two groups of Milky Way 
clusters identified by \citet{pfalzner09} are labeled.
\label{fig:age_size}}
\end{figure}

\begin{figure}
\plotone{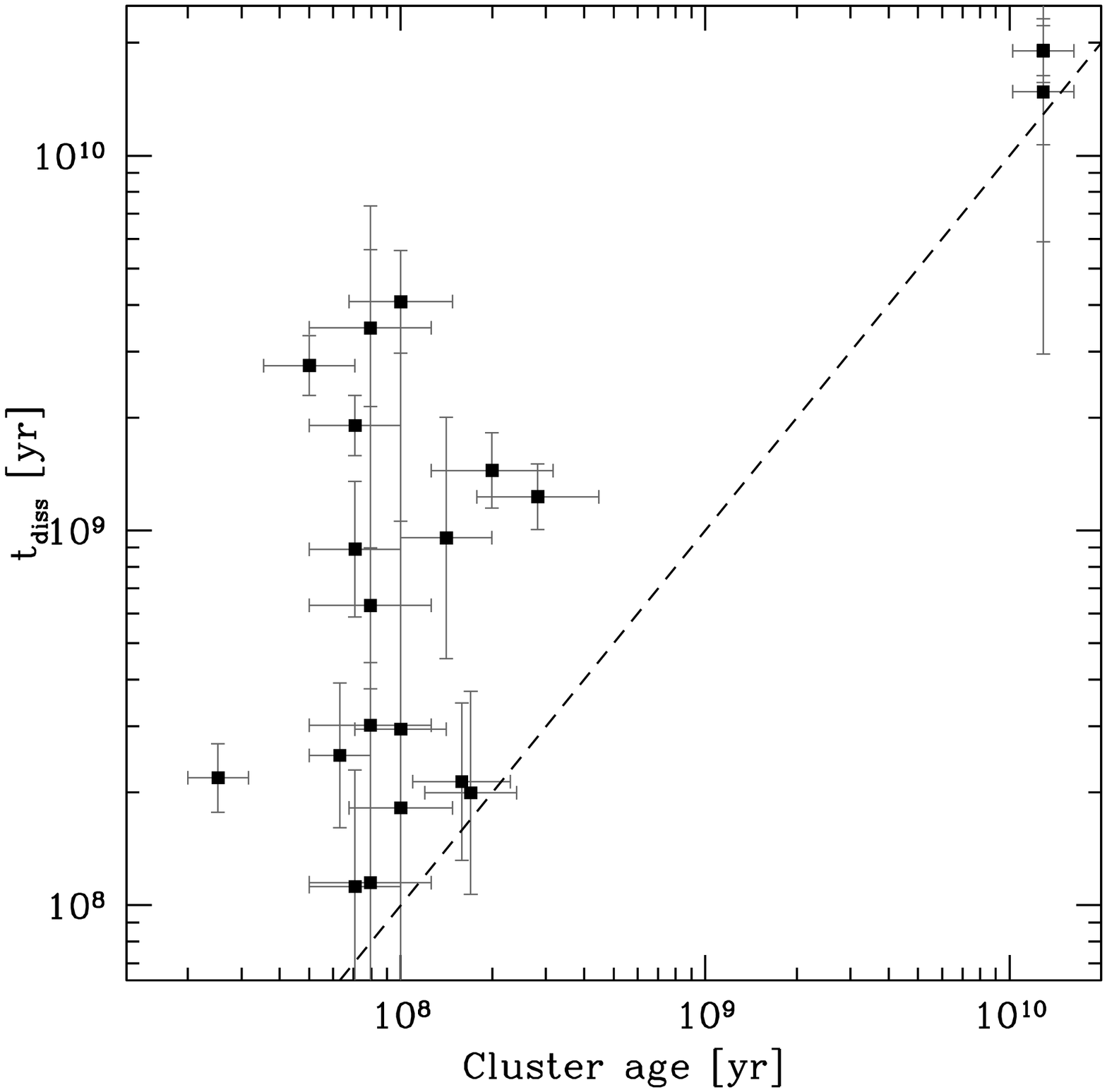}
\caption{
Dissolution times for M31 star clusters,  compared to cluster ages.
Four of the five old clusters are plotted at the same position, with 
dissolution times $20$~Gyr and assumed ages 13~Gyr.
\label{fig:t_dissolve}}
\end{figure}

\begin{figure}
\plotone{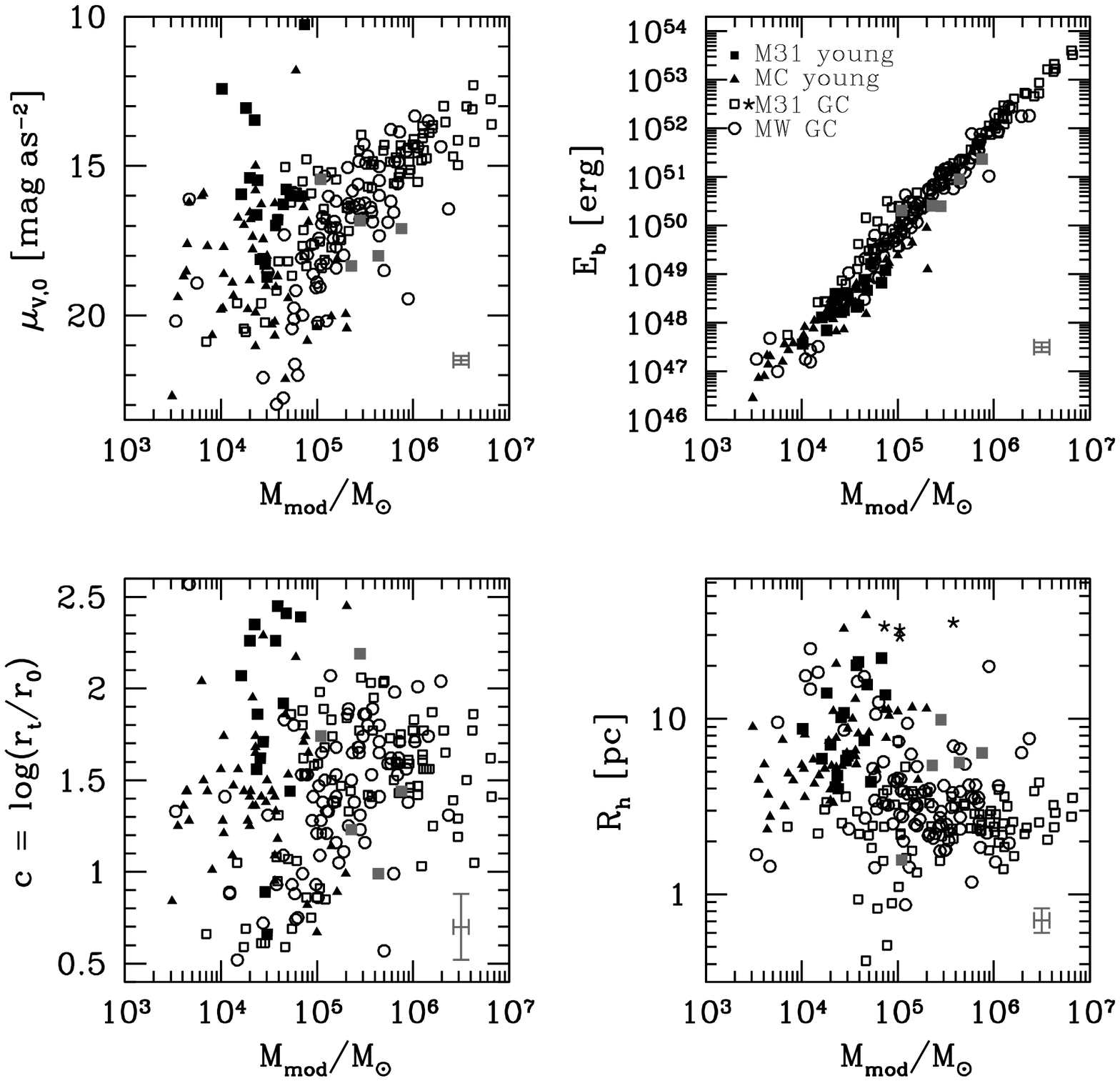}
\caption{
Structural properties of young and old star clusters in M31, young clusters
in the Magellanic Clouds, and globular clusters in the Milky Way,
shown as a function of cluster mass. 
Top left: central surface brightness; 
top right: binding energy;
lower left: concentration;
lower right: half-light radius.
Filled squares:  M31 clusters from the present sample
(black: young clusters, grey: old clusters); 
open squares: old M31 clusters from  \citet{barmby07,barmby02};
stars: `extended luminous clusters' in M31 from \citet{huxor05};
filled triangles: young Magellanic Cloud clusters.
Error bars show median uncertainties for the young M31 clusters.
\label{fig:struct_comp}}
\end{figure}

\begin{figure}
\plotone{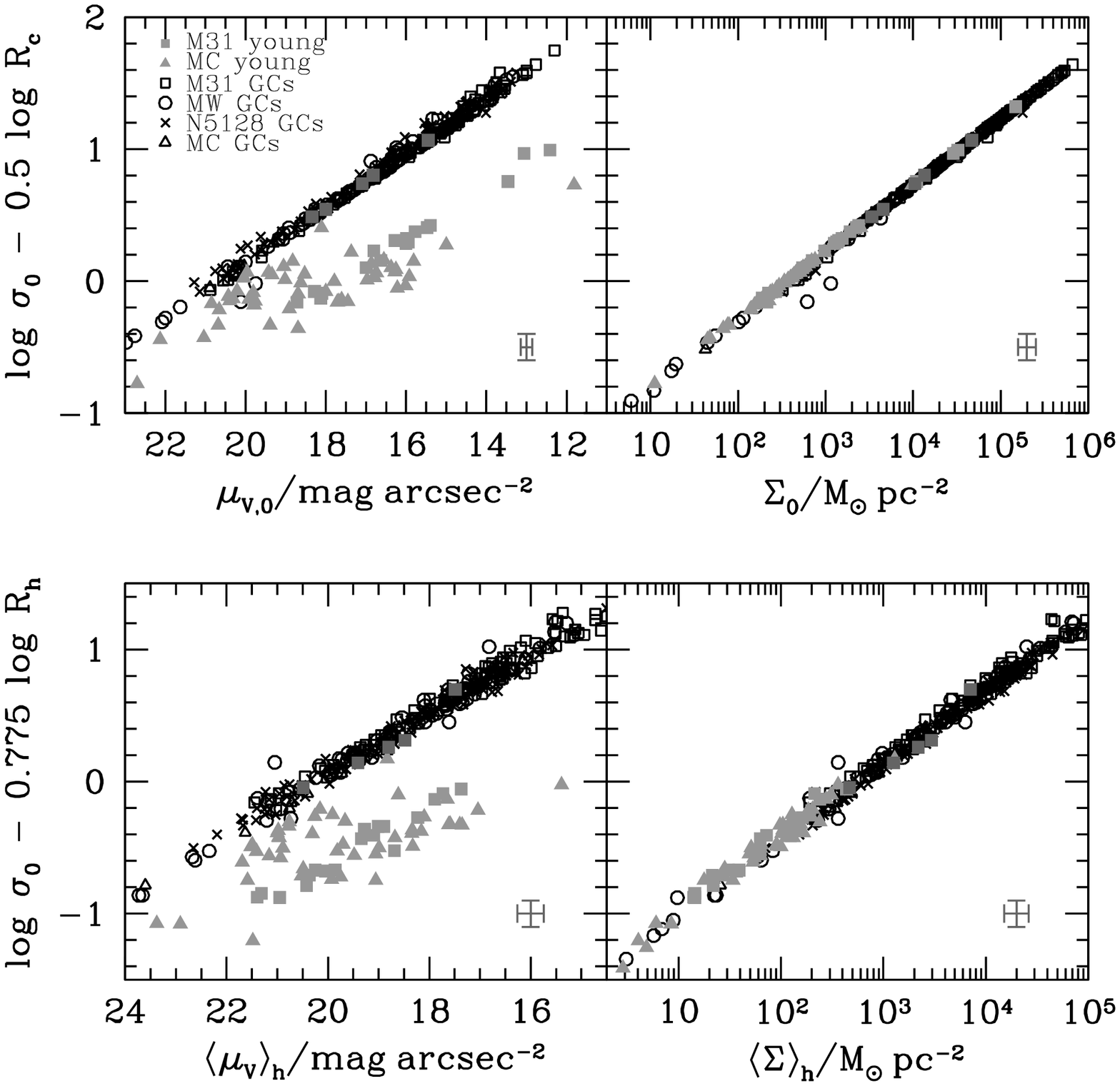}
\caption{
Views of the star cluster fundamental plane,
with core parameter relations in the bottom
panels and half-light parameter relations in
the top panels.
$\sigma_0$ is predicted central velocity dispersion 
and $\Sigma$ represents surface mass density 
either in the cluster core or at the half-light radius.
Left panels show surface brightness while right panels
show mass surface density.
Filled squares:  M31 clusters from the present sample
(light grey: young clusters, dark grey: old clusters); 
open triangles: old Magellanic Cloud and Fornax clusters; 
open circles: Milky Way globulars;
crosses:  NGC~5128 globulars. 
Other symbols as in Figure~\ref{fig:struct_comp}.
Error bars show median uncertainties for the young M31 clusters.
\label{fig:cluster_fp}}
\end{figure}

\begin{figure}
\plotone{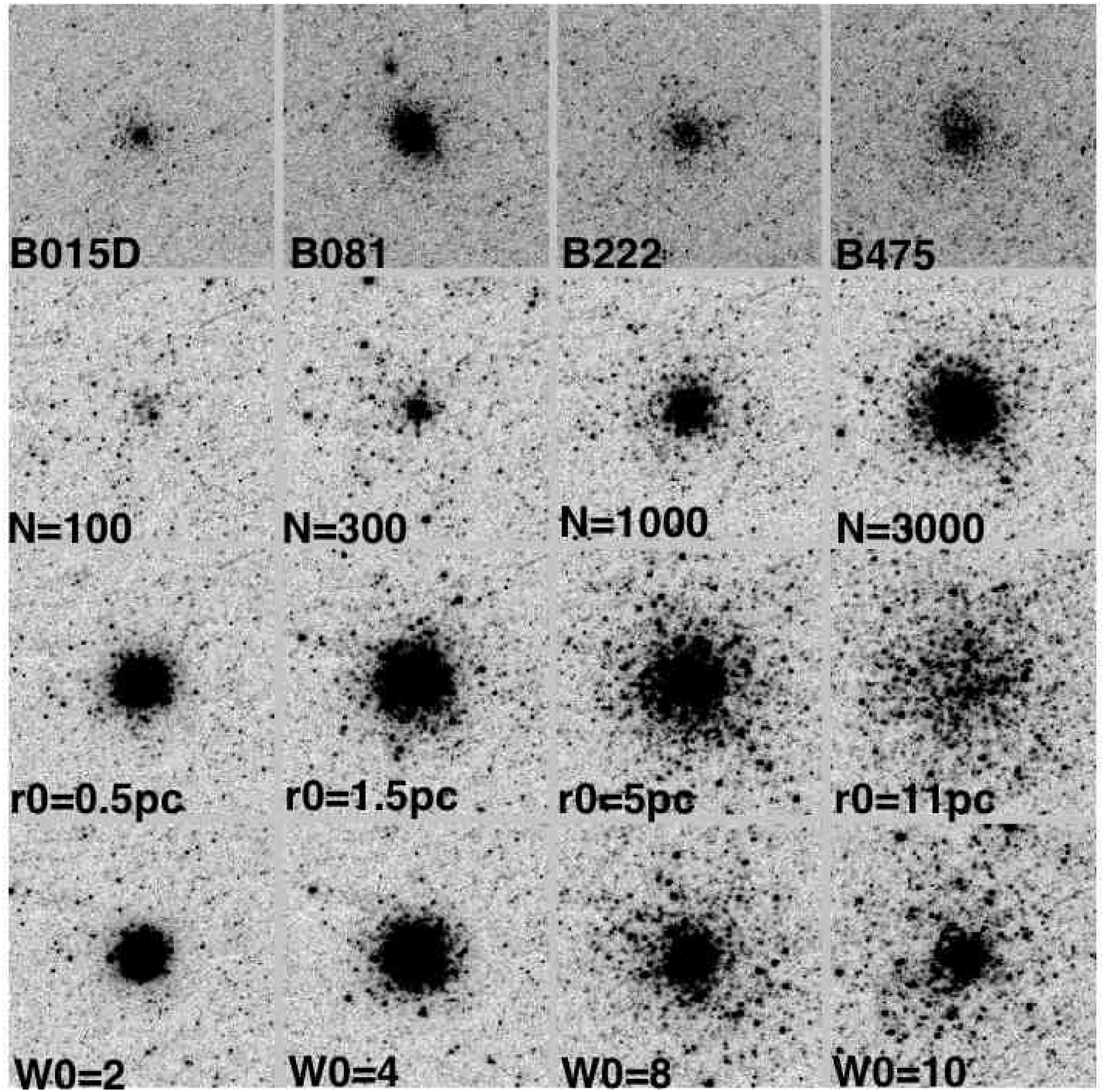}
\caption{
Top row: four M31 star clusters observed as
part of program GO-10818 with HST. Left to right: B015D, B081, B222, B475.
Second row: simulated clusters with central potential $W_0 = 6$
and scale radius $r_0=2$~pc, with (left to right) $N_*=100,300,1000,3000$.
Third row: simulated clusters with $N_*=3000$, central potential $W_0 = 6 $
and scale radius (left to right) $r_0=0.5,1.5,5,11$~pc.
Fourth row: simulated clusters with $N_*=3000$, scale radius $r_0=2$~pc,
and (left to right) $W_0 = 2,4,8,10$.
All images are 800~s exposures in the F450W filter on the WFPC2/PC chip;
each sub-image is  $13.7\times 13.7$ arcsec ($51.7\times 51.7$~pc at the 
distance of M31).
\label{fig:pix_f450}}
\end{figure}

\begin{figure}
\plotone{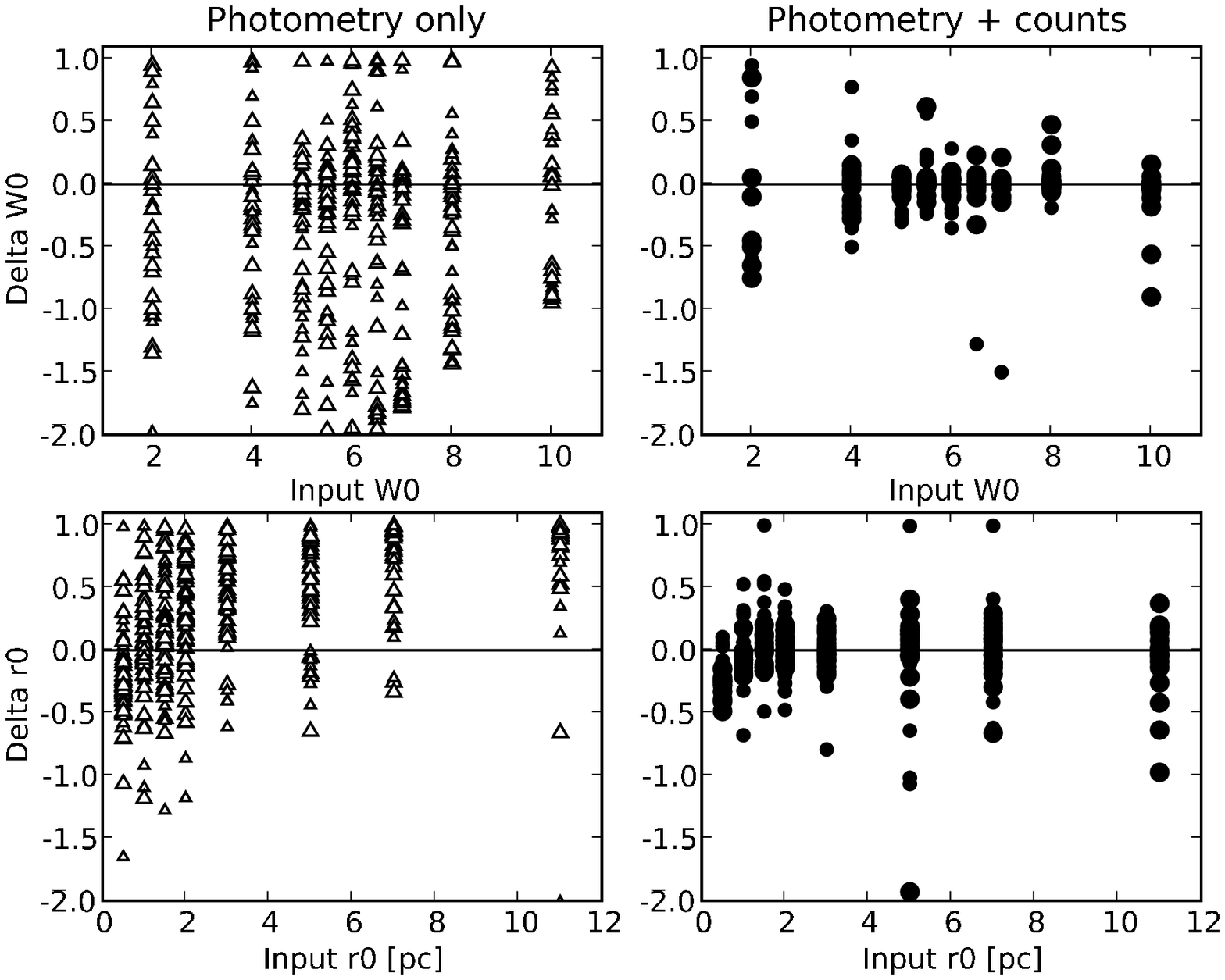}
\caption{Comparison of input and output structural parameters
for simulated star clusters. The ouput parameters are derived from
fitting Wilson models to surface density profiles derived from
simulated HST/WFPC2 images of the clusters.
Left: profiles measured with integrated photometry only;
right: profiles measured with integrated photometry  and number counts;
top: difference in central potential $\Delta W_0 = (W_{\rm 0,in} -W_{\rm 0,out})/W_{\rm 0,in} $;
bottom: difference in scale radius $\Delta r_0 = (r_{\rm 0,in} -r_{\rm 0,out})/r_{\rm 0,in} $;
\label{fig:comp_artphot}}
\end{figure}

\begin{figure}
\plotone{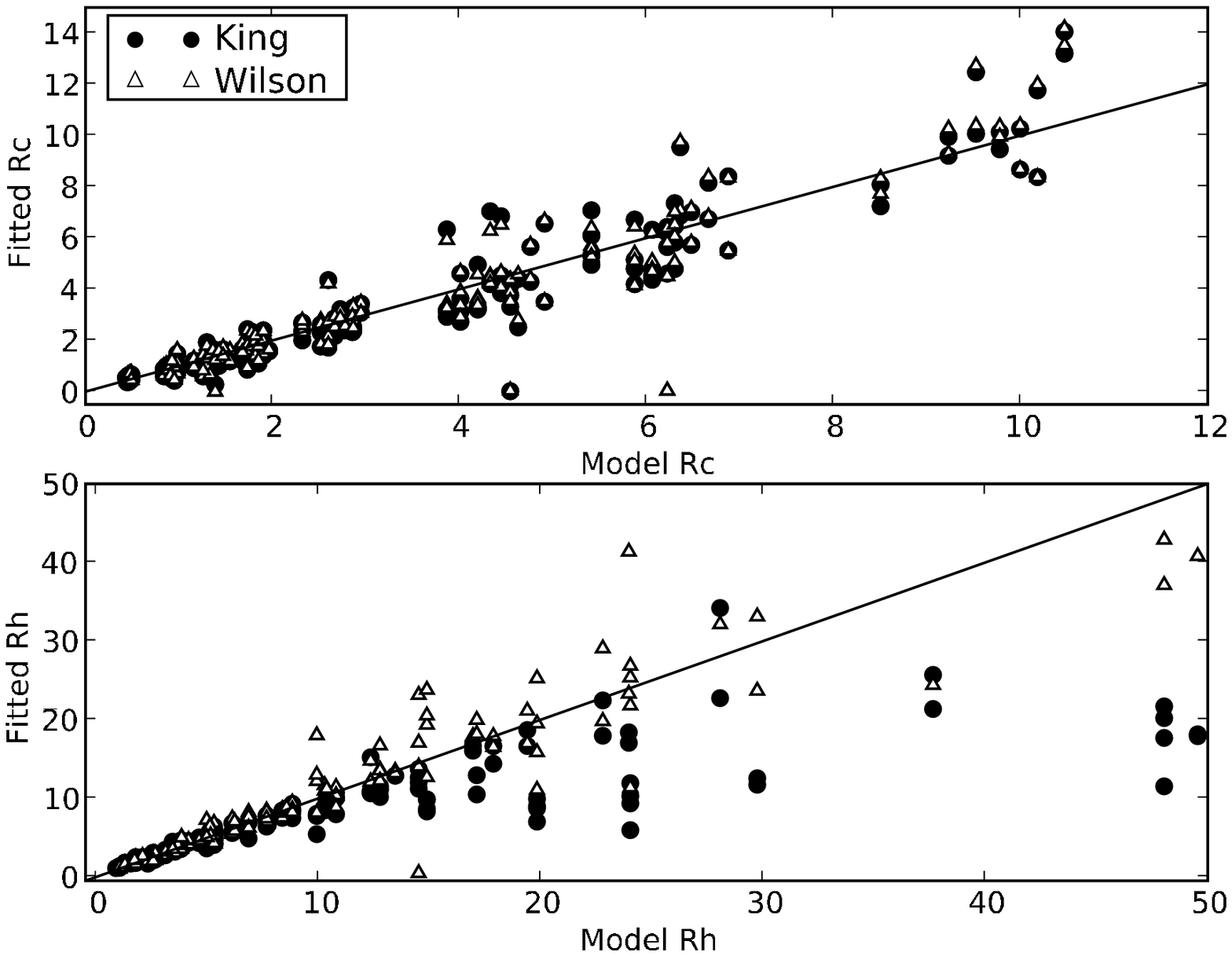}
\caption{Comparison of cluster size measurements
for fits of model density profiles to  artificial cluster profiles.
Top:  core radius  $R_c$; 
bottom: half-light radius $R_h$;
circles:  \citet{king66} model fits; 
triangles: \citet{wilson75} model fits.
\label{fig:art_king_wils}}
\end{figure}

\begin{figure}
\plotone{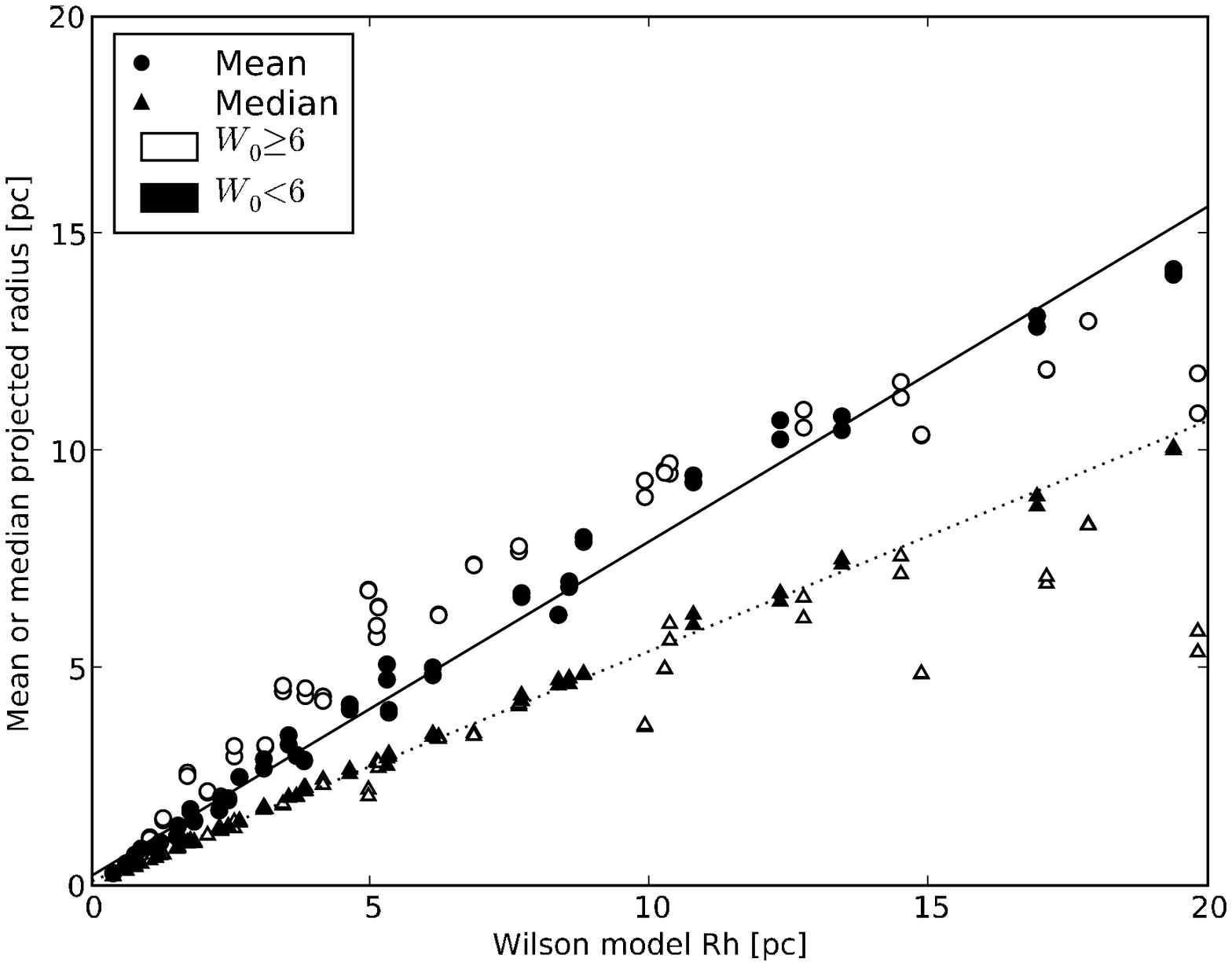}
\caption{Comparison of model half-light radius $R_h$ 
to mean and median projected radius for artificial clusters. 
Circles: mean; 
triangles: median;
filled symbols: models with $W_0<6$; 
open symbols: models with $W_0\geq 6$.
Solid line: least-squares fit to filled circles;
dotted line: least-squares fit to filled triangles.
\label{fig:mean_rh}}
\end{figure}

\clearpage

\begin{deluxetable}{lllllllll}
\tabletypesize{\scriptsize}
\tablecaption{Data for M31 young clusters\label{tab:hstdat}}
\tablewidth{0pt}
\tablehead{
\colhead{Name\tablenotemark{a}}  & \colhead{Dataset 1} & \colhead{Dataset 2} 
& \colhead{Filter 1} &\colhead{Exposure 1 [s]} & \colhead{Filter 2} &\colhead{Exposure 2 [s]}
& \colhead{$E(B-V)$} & \colhead{log age [yr]}}
\startdata
B015D     &u9pi140[12]  &u9pi140[34]&F450W& 800 &F814W& 800&0.65 &7.85\\  
B040	  &u9pi050[12]  &u9pi050[34]&F450W& 800 &F814W& 800&0.23 &7.90\\  
B043	  &u9pi022[12]  &u9pi022[34]&F450W& 800 &F814W& 800&0.23 &7.90\\  
B066	  &u9pi240[12]  &u9pi240[34]&F450W& 800 &F814W& 800&0.23 &7.85\\  
B081	  &u9pi170[12]  &u9pi170[34]&F450W& 800 &F814W& 800&0.30 &8.15\\  
B083      &u9pi250[12]  &u9pi250[34]&F450W& 800 &F814W& 800&0.20 &10.11\\  
B222	  &u9pi180[12]  &u9pi180[34]&F450W& 800 &F814W& 800&0.20 &8.90\\  
B257D     &u9pi100[12]  &u9pi100[34]&F450W& 800 &F814W& 800&0.30 &7.90\\  
B315	  &u5bj010[12]  &u5bj010[78]&F439W&1600 &F555W&1200&0.31 &8.00\\  
B318	  &u9pi020[12]  &u9pi020[34]&F450W& 800 &F814W& 800&0.17 &7.85\\  
B319	  &u5bj020[12]  &u5bj020[78]&F439W&1600 &F555W&1200&0.23 &8.00\\  
B321	  &u9pi150[12]  &u9pi150[34]&F450W& 800 &F814W& 800&0.25 &8.23\\  
B327	  &u9pi030[12]  &u9pi030[34]&F450W& 800 &F814W& 800&0.20 &7.70\\  
B342	  &u5bj030[12]  &u5bj030[78]&F439W&1600 &F555W&1200&0.20 &8.20\\  
B347      &u9pi230[12]  &u9pi230[34]&F450W& 800 &F814W& 800&0.06 &10.11\\  
B368	  &u5bj040[12]  &u5bj040[78]&F439W&1600 &F555W&1200&0.20 &7.80\\  
B374	  &u9pi070[12]  &u9pi070[34]&F450W& 800 &F814W& 800&0.30 &8.80\\  
B376	  &u9pi080[12]  &u9pi080[34]&F450W& 800 &F814W& 800&0.30 &8.00\\  
B448	  &u9pi200[12]  &u9pi200[34]&F450W& 800 &F814W& 800&0.35 &7.90\\  
B475	  &u9pi090[12]  &u9pi090[34]&F450W& 800 &F814W& 800&0.35 &8.30\\  
NB16      &u9pi120[12]  &u9pi012[34]&F450W& 800 &F814W& 800&0.25 &10.11\\   
V031      &u9pi130[12]  &u9pi130[34]&F450W& 800 &F814W& 800&0.35 &8.45\\  
VDB0	  &u9pi010[12]  &u9pi010[34]&F450W& 800 &F814W& 800&0.20 &7.60\\  
\enddata
\tablenotetext{a}{Naming convention of the Revised Bologna Catalog \citep{rbc04} is used. See that work for coordinates.}
\end{deluxetable}

\begin{deluxetable}{llll}
\tabletypesize{\scriptsize}
\tablecaption{Calibration data for WFPC2 imaging\label{tab:caldat}}
\tablewidth{0pt}
\tablehead{
\colhead{filter} & \colhead{zeropoint} & \colhead{$M_\sun$} &\colhead{Conversion factor\tablenotemark{a}}}
\startdata
F439W & 22.987& 5.55 &45.138\\
F450W & 23.996 & 5.31 &14.274\\
F555W & 24.621 & 4.83 &5.163\\
F814W & 23.641 & 4.14 &6.744 \\
\enddata
\tablenotetext{a}{Multiplicative conversion between surface brightness in counts~s$^{-1}$~arcsecc$^{-2}$
and intensity in L$_\sun$~pc$^{-2}$.}
\end{deluxetable}

\begin{deluxetable}{lccrrrrrrrr}
\tabletypesize{\scriptsize}
\rotate
\tablewidth{0pt}
\tablecaption{Basic Parameters of Fits to Profiles of M31 Young Clusters
\label{tab:fits}}
\tablehead{
\colhead{Name} & \colhead{Filter} &
\colhead{$N_{\rm pts}$} & \colhead{Model} & \colhead{$\chi_{\rm min}^2$}   &
\colhead{$I_{\rm bkg}$} & \colhead{$W_0$} & \colhead{$c$}         &
\colhead{$\mu_0$} & \colhead{$\log\,r_0$} & \colhead{$\log\,r_0$}         \\
\colhead{} & \colhead{}  & \colhead{}   &
\colhead{} & \colhead{} & \colhead{[$L_\odot\,{\rm pc}^{-2}$]} & \colhead{}  &
\colhead{} & \colhead{[mag arcsec$^{-2}$]} & \colhead{[arcsec]}              &
\colhead{[pc]} }
\startdata
B015D  & F450W        & $21$        & K66  & $323.12$  & $7.5$  & $10.20^{+0.90}_{-0.80}$  & $2.39^{+0.18}_{-0.17}$  & $16.12^{+0.15}_{-0.15}$  & $-0.640^{+0.108}_{-0.112}$  & $-0.061^{+0.108}_{-0.112}$ \\
          ~~              & ~~        & ~~         & W  & $386.35$  & $7.5$  & $10.80^{+1.10}_{-1.00}$  & $3.38^{+0.13}_{-0.05}$  & $16.11^{+0.16}_{-0.14}$  & $-0.650^{+0.121}_{-0.111}$  & $-0.071^{+0.121}_{-0.111}$ \\
          
B015D  & F814W          & $21$   & K66   & $231.70$  & $12.8$  & $14.40^{+1.40}_{-1.00}$  & $3.23^{+0.31}_{-0.21}$  & $12.61^{+0.48}_{-0.69}$  & $-1.758^{+0.196}_{-0.279}$  & $-1.179^{+0.196}_{-0.279}$ \\
  ~~        & ~~               & ~~    & W  &$ 377.92$  & $12.8$  & $14.90^{+1.50}_{-1.20}$  & $4.15^{+0.39}_{-0.30}$  & $12.47^{+0.51}_{-0.70}$  & $-1.804^{+0.215}_{-0.287}$  & $-1.225^{+0.215}_{-0.287}$ \\
  B040  & F450W          & $21$        & K66   & $44.18$  & $33.18\pm3.56$  & $9.60^{+0.40}_{-0.30}$  & $2.26^{+0.09}_{-0.07}$  & $15.44^{+0.08}_{-0.11}$  & $-0.967^{+0.048}_{-0.067}$  & $-0.387^{+0.048}_{-0.067}$ \\
                                     ~~        & ~~             & ~~          & W   & $50.75$  & $21.84\pm5.10$  & $9.80^{+0.50}_{-0.40}$  & $3.32^{+0.02}_{-0.00}$  & $15.48^{+0.08}_{-0.10}$  & $-0.931^{+0.054}_{-0.069}$  & $-0.352^{+0.054}_{-0.069}$ \\
\enddata
\tablecomments{
Table \ref{tab:fits} is available in its entirety in the electronic 
edition of the Journal. A short extract from it is shown here, for guidance
regarding its form and content. Column descriptions:
$\chi_{\rm min}^2$: unreduced $\chi^2$ of best-fitting model;
$I_{\rm bkg}$: model-fit background intensity (values without uncertainties indicate 
clusters for which the background was fixed manually);
$W_0$: model-fit central potential;
$c = \log(r_t/r_0)$: model-fit concentration ($r_t$ is tidal radius, given in Table~\ref{tab:mod_param});
$\mu_0$: model-fit central surface brightness;
$\log\,r_0$: model-fit scale radius.
Uncertainties are 68\% confidence intervals, computed as described in the text.
  }
\end{deluxetable}

\begin{deluxetable}{lccrrrrrrrrrr}
\tabletypesize{\scriptsize}
\rotate
\tablewidth{0pt}
\tablecaption{Derived Structural and Photometric Parameters for M31 Young Clusters
\label{tab:mod_param}}
\tablehead{
\colhead{Name} & \colhead{Filter} & \colhead{$V$ color} & \colhead{Model} &
\colhead{$\log\,r_{\rm tid}$} & \colhead{$\log\,R_c$} &
\colhead{$\log\,R_h$} & \colhead{$\log\,(R_h/R_c)$}   &
\colhead{$\log\,I_0$}  & \colhead{$\log\,j_0$}  & \colhead{$\log\,L_{V}$}  &
\colhead{$V_{\rm tot}$} & \colhead{$\log\,I_h$} \\
\colhead{} & \colhead{} & \colhead{[mag]} & \colhead{} & \colhead{[pc]} & \colhead{[pc]} &
\colhead{[pc]} & \colhead{} &
\colhead{[$L_{\odot, V}\,{\rm pc}^{-2}$]}    &
\colhead{[$L_{\odot, V}\,{\rm pc}^{-3}$]}    &
\colhead{[$L_{\odot, V}$]} & \colhead{[mag]} &
\colhead{[$L_{\odot, V}\,{\rm pc}^{-2}$]} 
}
\startdata
B015D  & F450W     &  $-0.114\pm0.1$   & K66   &$2.33^{+0.06}_{-0.07}$  & $-0.065^{+0.106}_{-0.110}$  & $1.346^{+0.100}_{-0.120}$  & $1.411^{+0.210}_{-0.226}$  & $4.16^{+0.07}_{-0.07}$  & $3.92^{+0.17}_{-0.17}$  & $5.89^{+0.06}_{-0.06}$  & $14.59^{+0.15}_{-0.16}$  & $2.39^{+0.20}_{-0.16}$ \\
          ~~        & ~~        & ~~       & W  & $3.30^{+0.07}_{-0.00}$  & $-0.076^{+0.118}_{-0.108}$  & $1.746^{+0.061}_{-0.051}$  & $1.821^{+0.170}_{-0.169}$  & $4.16^{+0.07}_{-0.08}$  & $3.93^{+0.17}_{-0.27}$  & $6.12^{+0.07}_{-0.05}$  & $13.99^{+0.14}_{-0.17}$  & $1.83^{+0.08}_{-0.08}$ \\
B015D  & F814W      & $0.457\pm0.1$  & K66  & $2.05^{+0.03}_{-0.01}$  & $-1.178^{+0.196}_{-0.279}$  & $1.086^{+0.014}_{-0.001}$  & $2.264^{+0.288}_{-0.194}$  & $5.33^{+0.28}_{-0.20}$  & $6.21^{+0.56}_{-0.39}$  & $5.75^{+0.04}_{-0.04}$  & $14.93^{+0.10}_{-0.11}$  & $2.78^{+0.04}_{-0.04}$ \\

                                  ~~        & ~~   & ~~        & W    & $2.93^{+0.10}_{-0.08}$  & $-1.224^{+0.215}_{-0.286}$  & $1.312^{+0.053}_{-0.025}$  & $2.537^{+0.340}_{-0.240}$  & $5.39^{+0.28}_{-0.21}$  & $6.31^{+0.57}_{-0.42}$  & $5.87^{+0.05}_{-0.05}$  & $14.61^{+0.12}_{-0.13}$  & $2.45^{+0.05}_{-0.09}$ \\

                                  B040  & F450W     & $-0.029\pm0.1$   & K66   & $1.88^{+0.02}_{-0.02}$  & $-0.393^{+0.047}_{-0.066}$  & $0.853^{+0.047}_{-0.045}$  & $1.245^{+0.113}_{-0.092}$  & $4.40^{+0.06}_{-0.05}$  & $4.49^{+0.12}_{-0.09}$  & $5.33^{+0.04}_{-0.04}$  & $15.98^{+0.10}_{-0.10}$  & $2.82^{+0.09}_{-0.09}$ \\
                                   ~~        & ~~     & ~~           & W   & $2.97^{+0.05}_{-0.05}$  & $-0.361^{+0.052}_{-0.067}$  & $1.292^{+0.022}_{-0.032}$  & $1.652^{+0.089}_{-0.084}$  & $4.39^{+0.06}_{-0.05}$  & $4.54^{+0.04}_{-0.17}$  & $5.57^{+0.04}_{-0.04}$  & $15.37^{+0.11}_{-0.10}$  & $2.19^{+0.06}_{-0.05}$ \\
          \enddata
\tablecomments{
Table \ref{tab:mod_param} is available in its entirety in the
electronic  edition of the Journal. A short extract from it is shown here, for guidance
regarding its form and content. Column descriptions:
$r_t$: model tidal radius ($\rho(r_t)=0$); 
$R_c$: model projected  core radius, at which intensity is half the central value;
$R_h$: model projected half-light, or effective, radius (contains half the total luminosity in projection);
$R_h/R_c$: measure of cluster concentration;
$I_0$: model central luminosity surface density in the $V$ band;
$j_0$: logarithmic central luminosity volume density in the $V$ band;
$L_V$: total integrated model luminosity in the $V$ band;
$V_{\rm tot}=4.83-2.5\,\log\,(L_V/L_\odot)+ 5\,\log\,(D/10\,{\rm pc})$:
total, {\it extinction-corrected} apparent $V$-band magnitude;
$I_h \equiv L_V/2\pi R_h^2$: $V$-band
luminosity surface density averaged over the half-light radius.
Uncertainties are 68\% confidence intervals, computed as described in the text.
}

\end{deluxetable}

\begin{deluxetable}{lccrrrrrrrrrr}
\tabletypesize{\scriptsize}
\rotate
\tablewidth{0pt}
\tablecaption{Derived Dynamical Parameters  for M31 Young Clusters
\label{tab:mod_dynam}}
\tablecolumns{13}
\tablehead{
\colhead{Name} & \colhead{Filter} & \colhead{$\Upsilon_V^{\rm pop}$}     &
\colhead{Model} & \colhead{$\log\,M_{\rm tot}$} & \colhead{$\log\,E_b$}  &
\colhead{$\log\,\Sigma_0$} & \colhead{$\log\,\rho_0$} & 
\colhead{$\log\,\Sigma_{\rm h}$} & \colhead{$\log\,\sigma_{{\rm p},0}$}     &
\colhead{$\log\,v_{{\rm esc},0}$} & \colhead{$\log\,t_{\rm rh}$} &
\colhead{$\log\,f_0$}                                            \\
\colhead{} & \colhead{} & \colhead{[$M_\odot\,L_{\odot,V}^{-1}$]}  &
\colhead{} & \colhead{[$M_\odot$]} & \colhead{[erg]} &
\colhead{[$M_\odot$ pc$^{-2}$]} & \colhead{[$M_\odot$ pc$^{-3}$]} &
\colhead{[$M_\odot$ pc$^{-2}$]} & \colhead{[km~s$^{-1}$]} & \colhead{[km~s$^{-1}$]} &
\colhead{[yr]} & \colhead{[$M_\odot$ (pc km~s$^{-1}$)$^{-3}$]} }
\startdata
B015D & F450W  & $0.088^{+0.01}_{-0.01}$       & K66   & $4.83^{+0.08}_{-0.08}$  & $48.82^{+0.09}_{-0.09}$  & $3.10^{+0.09}_{-0.09}$  & $2.86^{+0.18}_{-0.18}$  & $1.34^{+0.21}_{-0.16}$  & $0.256^{+0.039}_{-0.042}$  & $0.914^{+0.032}_{-0.034}$  & $9.91^{+0.17}_{-0.20}$  & $0.891^{+0.251}_{-0.242}$ \\
          ~~        & ~~        & ~~         & W  & $5.07^{+0.08}_{-0.08}$  & $48.91^{+4.13}_{-3.46}$  & $3.11^{+0.08}_{-0.09}$  & $2.88^{+0.18}_{-0.27}$  & $0.78^{+0.09}_{-0.09}$  & $0.251^{+0.076}_{-0.043}$  & $0.924^{+0.548}_{-0.034}$  & $10.61^{+0.12}_{-0.10}$  & $0.915^{+0.250}_{-0.351}$ \\
  B015D  & F814W  & $0.088^{+0.01}_{-0.01}$    & K66  & $4.69^{+0.06}_{-0.07}$  & $48.76^{+0.09}_{-0.09}$  & $4.28^{+0.28}_{-0.20}$  & $5.15^{+0.56}_{-0.39}$  & $1.72^{+0.06}_{-0.07}$  & $0.286^{+0.031}_{-0.033}$  & $1.017^{+0.036}_{-0.036}$  & $9.47^{+0.04}_{-0.03}$  & $3.096^{+0.561}_{-0.395}$ \\
            ~~        & ~~        & ~~        & W     & $4.82^{+0.07}_{-0.07}$  & $48.80^{+0.09}_{-0.10}$  & $4.33^{+0.29}_{-0.22}$  & $5.25^{+0.57}_{-0.43}$  & $1.40^{+0.07}_{-0.10}$  & $0.290^{+0.031}_{-0.033}$  & $1.028^{+0.035}_{-0.036}$  & $9.87^{+0.10}_{-0.06}$  & $3.184^{+0.578}_{-0.437}$ \\
B040  & F450W  & $0.094^{+0.01}_{-0.01}$       & K66   & $4.30^{+0.06}_{-0.06}$  & $48.25^{+0.09}_{-0.09}$  & $3.38^{+0.07}_{-0.07}$  & $3.46^{+0.12}_{-0.10}$  & $1.80^{+0.10}_{-0.11}$  & $0.229^{+0.031}_{-0.034}$  & $0.875^{+0.030}_{-0.032}$  & $8.94^{+0.08}_{-0.08}$  & $1.570^{+0.148}_{-0.109}$ \\
      ~~        & ~~        & ~~         & W  & $4.54^{+0.06}_{-0.06}$  & $44.89^{+7.58}_{-0.09}$  & $3.36^{+0.07}_{-0.07}$  & $3.51^{+0.06}_{-0.18}$  & $1.16^{+0.08}_{-0.07}$  & $0.198^{+0.087}_{-0.032}$  & $0.940^{+0.530}_{-0.062}$  & $9.70^{+0.05}_{-0.06}$  & $1.604^{+0.050}_{-0.183}$ \\
                                          
\enddata
\tablecomments{
Table \ref{tab:mod_dynam} is available in its entirety in the electronic 
edition of the Journal. A short extract from it is shown here, for guidance
regarding its form and content. Column descriptions:
$\Upsilon_V^{\rm pop} L_V$: assumed mass-to-light ratio in the $V$ band;
$M_{\rm tot}=\Upsilon_V^{\rm pop} L_V$: integrated model mass; 
$E_b \equiv -(1/2)\int_{0}^{r_t} 4\pi r^2 \rho \phi\,dr$:  integrated binding energy;
$\Sigma_0$: central surface mass density;
${\rho}_0$: central volume density;
$\Sigma_h$: surface mass density averaged over  the half-light radius;
$\sigma_{{\rm p},0}$: predicted line-of-sight velocity dispersion at cluster center;
$v_{{\rm esc},0}$: predicted central ``escape'' velocity;
$\log\,t_{\rm rh}$: two-body relaxation time at model projected half-mass radius;
$\log f_0\equiv \log\,\left[\rho_0/(2\pi \sigma_c^2)^{3/2}\right]$:
a measure of the model's central  phase-space density or relaxation time.
For $f_0$ in these units, and $t_{rc}$ in years, $\log t_{rc} \simeq 8.28 - \log f_0$ 
\citep{mclaughlin05}.
Uncertainties are 68\% confidence intervals, computed as described in the text.
}
\end{deluxetable}

\end{document}